\documentclass{aa}  

\usepackage{graphicx}
\usepackage{txfonts}
\usepackage{mathtools}
\usepackage{soul}
\usepackage{url}
\urldef{\footurl}\url{https://www.sdss.org/dr17/mastar/mastar-stellar-parameters/}
\usepackage{color}

\begin{document}

   \title{SDSS-IV MaStar: Stellar parameter determination with continuum-supplemented full-spectrum fitting}

  \author{Daniel Lazarz\inst{1}\fnmsep\thanks{E-mail: dan.lazarz.mail@gmail.com}
          \and Renbin Yan\inst{2}\fnmsep\thanks{E-mail: rbyan@cuhk.edu.hk}
          \and Ronald Wilhelm\inst{1}
          \and Yanping Chen\inst{3}
          \and Lewis Hill\inst{4}
          \and Jon A. Holtzman\inst{5}
          \and Julie Imig\inst{5}
          \and Claudia Maraston\inst{6}
          \and Szabolcs Mészáros\inst{7,8}
          \and Guy S. Stringfellow\inst{9}
          \and Daniel Thomas\inst{6}
          \and Timothy C. Beers\inst{10}
          \and Dmitry Bizyaev\inst{11,12}
          \and Niv Drory\inst{13}
          \and Richard R. Lane\inst{14}
          \and Christian Nitschelm\inst{15}
          }

   \institute{Department of Physics and Astronomy, University of Kentucky, 505 Rose St., Lexington, KY 40506-0057, USA \\
        \and
             Department of Physics,The Chinese University of Hong Kong, Shatin, N.T., Hong Kong S.A.R., People’s Republic of China \\
        \and
             New York University Abu Dhabi, Abu Dhabi, P.O. Box 129188, United Arab Emirates
        \and
             Institute of Cosmology and Gravitation, University of Portsmouth, Burnaby Road, Portsmouth PO1 3FX, UK
        \and
             Department of Astronomy, New Mexico State University, Box 30001, MSC 4500, Las Cruces NM 88003, USA
        \and
             Institute of Cosmology \& Gravitation, University of Portsmouth, Dennis Sciama Building, Portsmouth, PO1 3FX, UK
        \and
             ELTE E\"otv\"os Lor\'and University, Gothard Astrophysical Observatory, 9700 Szombathely, Szent Imre H. st. 112, Hungary; MTA-ELTE Exoplanet Research Group
        \and
             MTA-ELTE Lend{\"u}let "Momentum" Milky Way Research Group, Hungary
        \and
             Center for Astrophysics and Space Astronomy, Department of Astrophysical and Planetary Sciences, University of Colorado, 389 UCB, Boulder, CO 80309-0389, USA
        \and
             Department of Physics and Astronomy and JINA Center for the Evolution of the Elements (JINA-CEE), University of Notre Dame, Notre Dame, IN 46556 USA
        \and
             Apache Point Observatory and New Mexico State University, P.O. Box 59, Sunspot, NM 88349, USA
        \and
             Sternberg Astronomical Institute, Moscow State University, Universitetskij pr. 13, Moscow, Russia
        \and
             McDonald Observatory, The University of Texas at Austin, 1 University Station, Austin, TX 78712, USA
        \and
             Centro de Investigación en Astronomía, Universidad Bernardo O'Higgins, Avenida Viel 1497, Santiago, Chile
        \and
             Centro de Astronom{\'i}a (CITEVA), Universidad de Antofagasta, Avenida Angamos 601, Antofagasta 1270300, Chile
             }

 
  \abstract
   {}
   {We present a stellar parameter catalog built to accompany the MaStar Stellar Library, which is a comprehensive collection of empirical, medium-resolution stellar spectra.}
   {We constructed this parameter catalog by using a multicomponent $\chi^{2}$ fitting approach to match MaStar spectra to models generated by interpolating the ATLAS9-based BOSZ model spectra. The total $\chi^{2}$ for a given model is defined as the sum of components constructed to characterize narrow-band features of observed spectra (e.g., absorption lines) and the broadband continuum shape separately. Extinction and systematics due to flux calibration were taken into account in the fitting. The $\chi^{2}$ distribution for a given region of model space was sampled using a Markov Chain Monte Carlo (MCMC) algorithm, the data from which were then used to extract atmospheric parameter estimates ($T_{\rm eff}$, $\log g$, [Fe/H], and [$\alpha$/Fe]), their corresponding uncertainties, and direct extinction measurements.}
   {Two methods were used to extract parameters and uncertainties: one that accepts the MCMC's prescribed minimum-$\chi^{2}$ result, and one that uses Bayesian inference to compute a likelihood-weighted mean from the $\chi^{2}$ distribution sampled by the MCMC. Results were evaluated for internal consistency using repeat observations where available and by comparing them with external data sets (e.g., APOGEE-2 and Gaia DR2). Our spectral-fitting exercise reveals possible deficiencies in current theoretical model spectra, illustrating the potential power of MaStar spectra for helping to improve the models. This paper represents an update to the parameters that were originally presented with SDSS-IV DR17. The MaStar parameter catalog containing our BestFit results is available on the SDSS-IV DR17 website as part of version 2 of the MaStar stellar parameter value-added catalog.}
   {}

   \keywords{Stellar Library --
                Stellar Parameter Determination --
                Atmospheric Parameters -- Extinction
               }

   \maketitle
%

\section{Introduction}

\label{introduction}

    Libraries of stellar spectra represent an essential component of Stellar-population Synthesis (SPS), alongside stellar isochrone models and an initial mass function (IMF). They can be divided into two broad categories: theoretical and empirical spectral libraries. Theoretical (or synthetic) spectra come with a number of advantages. These include arbitrarily high spectral resolution and more comprehensive stellar parameter space coverage than modern observing techniques allow. However, they suffer from a number of drawbacks as well, including incomplete atomic and molecular line lists due to inadequacies in our present understanding of stellar atmospheric physics. Empirical stellar spectra refer to spectra obtained from observations of actual stars. While they are limited in spectral resolution and parameter space coverage by observation techniques and the availability of desirable targets, they come with the benefit of providing the most accurate possible representation of real-world stellar physics.
    
    Previous empirical stellar libraries have proven invaluable for stellar population research. Examples of such libraries include MILES \citep{Sanchez-Blazquez06,FalconBarroso11}, the X-Shooter Stellar Library (XSL, \citealt{Chen14}; \citealt{Verro_2022}), ELODIE \citep{Soubiran98, PrugnielS01, PrugnielS04, Prugniel07}, \cite{GunnS83}, Pickles \citep{Pickles85, Pickles98}, \cite{Diaz89}, \cite{SilvaC92}, Lick/IDS \citep{Worthey94}, \cite{LanconW00}, STELIB \citep{LeBorgne03}, INDO-US  \citep{Valdes04}, CaT \citep{Cenarro01}, HST NGSL \citep{Gregg06}, the NASA Infrared Telescope Facility (IRTF) Library \citep{Rayner09}, and the Extended IRTF library \citep{Villaume17}. However, these libraries have several significant shortcomings, including limited wavelength and parameter coverage. A project on the scale of Mapping Nearby Galaxies at Apache Point Observatory (MaNGA, \citealt{Bundy15,yan16b}) stands to benefit greatly from a larger library with a wider variety of stellar types of varying parameters. MaNGA is a large survey containing detailed integral field unit (IFU) spectroscopy data for approximately $10,000$ galaxies, and it was one of three major projects undertaken as part of the fourth generation of the Sloan Digital Sky Survey (SDSS-IV, \citealt{Blanton17}), alongside the extended Baryon Oscillation Spectroscopic Survey (eBOSS, \citealt{Dawson16}), and the second generation of the Apache Point Observatory Galactic Evolution Experiment (APOGEE-2, \citealt{Majewski16}).
    
    In an effort to address the issues with present stellar libraries, we have constructed the MaNGA stellar library (MaStar, \citealt{Yan_2019}), an empirical stellar library built to accompany MaNGA using the same instrumentation via parallel observing. The final library contains more than $25,000$ empirical spectra from nearly $12,000$ unique science targets, supplemented by more than $33,500$ spectra from more than $12,000$ unique standard stellar targets. These spectra are uniform in quality and provide a broad and comprehensive coverage of stellar parameter space. These parameters include effective temperature, surface gravity, iron abundance (as a proxy for overall stellar metallicity), and $\alpha$ abundance, hereafter $T_{\rm eff}$, $\log g$, [Fe/H], and [$\alpha$/Fe].
    
    Of the three aforementioned ingredients for SPS (the IMF, stellar isochrones, and stellar spectra), the last two are closely linked as they contain the information needed to understand how individual stellar atmospheres evolve. This information is necessary for reproducing composite spectra of stellar populations. However, in order to make use of isochrone data in conjunction with empirical stellar spectra, it is necessary to have a set of accurate stellar parameters at the correct locations along the stars' evolutionary tracks. The three atmospheric parameters, $T_{\rm eff}$, $\log g$, and [Fe/H], are usually sufficient for this task, but additional parameters, such as $\alpha$ abundance, are also useful for recreating abundance patterns observed in nature.
    
    There are several common methods for estimating stellar parameters, including some that involve the use of sophisticated machine learning algorithms, such as \emph{The Cannon} or \emph{The Payne} \citep{Ness15,Ting_2019}. These are data-driven approaches that require a training set of stellar spectra for which the desired parameters are assumed to already be known accurately. While these methods have the advantage of being primarily based on empirical spectral data, they are limited by the availability, quality, and parameter coverage of the training set being used. In other words, the set of estimated stellar parameters obtained is only as good as the input training set, both in accuracy and coverage. As a consequence, such data-driven methods have difficulty assigning parameters to a star that lies outside the region of parameter space spanned by its training set. While this approach is still extremely useful for many applications, it presents a problem when developing parameter sets for newer libraries that aim to obtain broader parameter coverage than previous catalogs, which is one of the primary goals of MaStar. As a result, such machine learning approaches should be treated with caution.
    
    Another category of methods involves matching theoretical templates with predetermined parameters to empirical spectra in order to find the best-fitting model, and hence, the best-fitting stellar parameters. Such model-based methods have the advantage of not being limited by the parameter space coverage or quality of previously obtained empirical spectra, since models can be generated for a wide variety of stars at whatever resolution is desired. However, this approach relies on our previously mentioned incomplete understanding of the stellar atmospheric physics. As a result, precautions must be taken, and the possibility that the models may contain subtle inaccuracies must be considered.
    
    In this paper, we present our set of parameters for the MaStar library and discuss our approach to parameter determination. Our method used a set of theoretical model spectra to perform a minimum-$\chi^{2}$ fitting of the data in order to obtain best-fitting models via interpolation according to several measures. These measures included comparing the data with models according to narrow-band spectral features, such as absorption lines, as well as broadband features that make up the continuum shape. Our method's use of the information contained in the continuum in addition to the continuum-normalized full-spectrum fitting makes it unique. Various complicating effects were taken into account, such as interstellar extinction, and imperfect flux calibration. The final product is a set of stellar parameters ($T_{\rm eff}$, $\log g$, [Fe/H], and [$\alpha$/Fe]) for the complete set of MaStar spectra, ready for use in stellar-population synthesis. Our independently derived extinction estimates, in the form of $A_V$, are also included with the final parameter set.
    
    There have been several other concurrent parameter determination efforts for MaStar underway in recent years. \cite{Hill2021} presents a method that uses a version of the penalized pixel-fitting method (pPXF, \citealt{CapEm2004}; \citealt{cap2016}) for full-spectrum fitting, using a Bayesian approach for parameter estimation with priors based on Gaia photometry \citep{gaia_2018}. \cite{Imig2022} presents a data-driven approach that uses a neural network model trained on previously obtained APOGEE-2 parameters for a subset of MaStar targets. Chen et al. (2022, in preparation) will present a Bayesian method using priors based on isochrone data. \cite{chen2020} also presents an early parameter set for MaStar's first data release as a part of SDSS DR15.
    
    In this paper, we discuss our approach to stellar parameter determination in detail in Section \ref{method}, covering each step of the procedure. In Section \ref{results}, we evaluate the reliability of the parameters through internal and external comparisons and discuss the implications of our results on the present state of atmospheric models. Finally, in Section \ref{conclusions}, we discuss our closing thoughts and the potential for future work.

\section{Method}
\label{method}

\subsection{Overview}
\label{overview}

    The method we developed combines the use of traditional continuum-normalized spectral fitting with a more novel form of continuum fitting. The goal of this was to use the information contained within a star's broadband shape to place constraints on the traditional spectral-fitting procedure, and break degeneracies that arise in certain regions of parameter space. This is particularly helpful for hotter stars, with temperatures above $~7,000$  K. The reasoning behind this is that the continuum itself characterizes certain useful parameter-dependent features (e.g., the Balmer break) in a similar way to photometry. So, in essence, this approach combines the spirit of spectroscopic parameter fitting with that of traditional photometric fitting to arrive at a solution that is consistent with both methods. 
    
    The procedure we used to compare a MaStar spectrum with a given model involved a reduced-$\chi^{2}$ calculation, in which a total $\chi^{2}$ was defined as the weighted sum of three components, defined as the following:
    
    \begin{enumerate}
        \item \textbf{High-Frequency Component ($\chi^{2}_{\rm HF}$)}: Computed by comparing the continuum-normalized MaStar spectrum and continuum-normalized model spectrum directly.
        
        \item \textbf{Low-Frequency Component ($\chi^{2}_{\rm LF}$)}: Computed by comparing the large-wavelength-scale (or ``low-frequency'') features of the MaStar spectrum with that of the model spectrum with foreground extinction applied.
        
        \item \textbf{Flux Calibration Component ($\chi^{2}_{\rm F}$)}: Computed by evaluating the polynomial correction needed to account for flux-calibration residuals. This generally amounted to only a small addition to the total $\chi^{2}$.
        
    \end{enumerate}
    
    \noindent{These three components were combined additively to give $\chi^{2}_{\rm Total}$, which was used to quantify a given model's overall agreement with the data. With $\chi^{2}_{\rm Total}$ defined, we were free to use whichever minimization algorithm we preferred to find the model that minimized $\chi^{2}_{\rm Total}$, and take that model's parameters to be correct. In practice, this procedure came with several caveats, which we discuss in later sections. These included the need for strategic weighting of certain components of $\chi^{2}_{\rm Total}$, as well as the masking or deweighting of certain spectral features that were believed to be inaccurately represented by the models.}

    The $\chi^{2}$ fitting was carried out in four phases:
     
    \begin{itemize}
        \item[] \textbf{Phase 1: Initialization --} Prepare the model set according to the given MaStar spectrum's unique instrumental line spread function (LSF), approximate the continua, and generate continuum-normalized spectra for both the data and the models using an identical procedure.
        
        \item[] \textbf{Phase 2: Global Grid Search --} Find the best-fitting discrete model grid point from the continuum-normalized spectral fitting to use as an initial parameter estimate for the next phase.
        
        \item[] \textbf{Phase 3: Interpolator Search --} Use an MCMC (Markov Chain Monte Carlo) algorithm to sample the distribution of $\chi^{2}_{\rm Total}$ in the vicinity of the initial estimate.
        
        \item[] \textbf{Phase 4: Parameter Extraction --} Derive parameter values and corresponding error estimates from the MCMC data.
    \end{itemize}
    
   \noindent{These phases were performed for each individual MaStar spectrum, and diagnostic data were recorded where appropriate for use in post processing.}
   
\begin{figure*}
        \centering
        \includegraphics[width=0.95\textwidth]
        {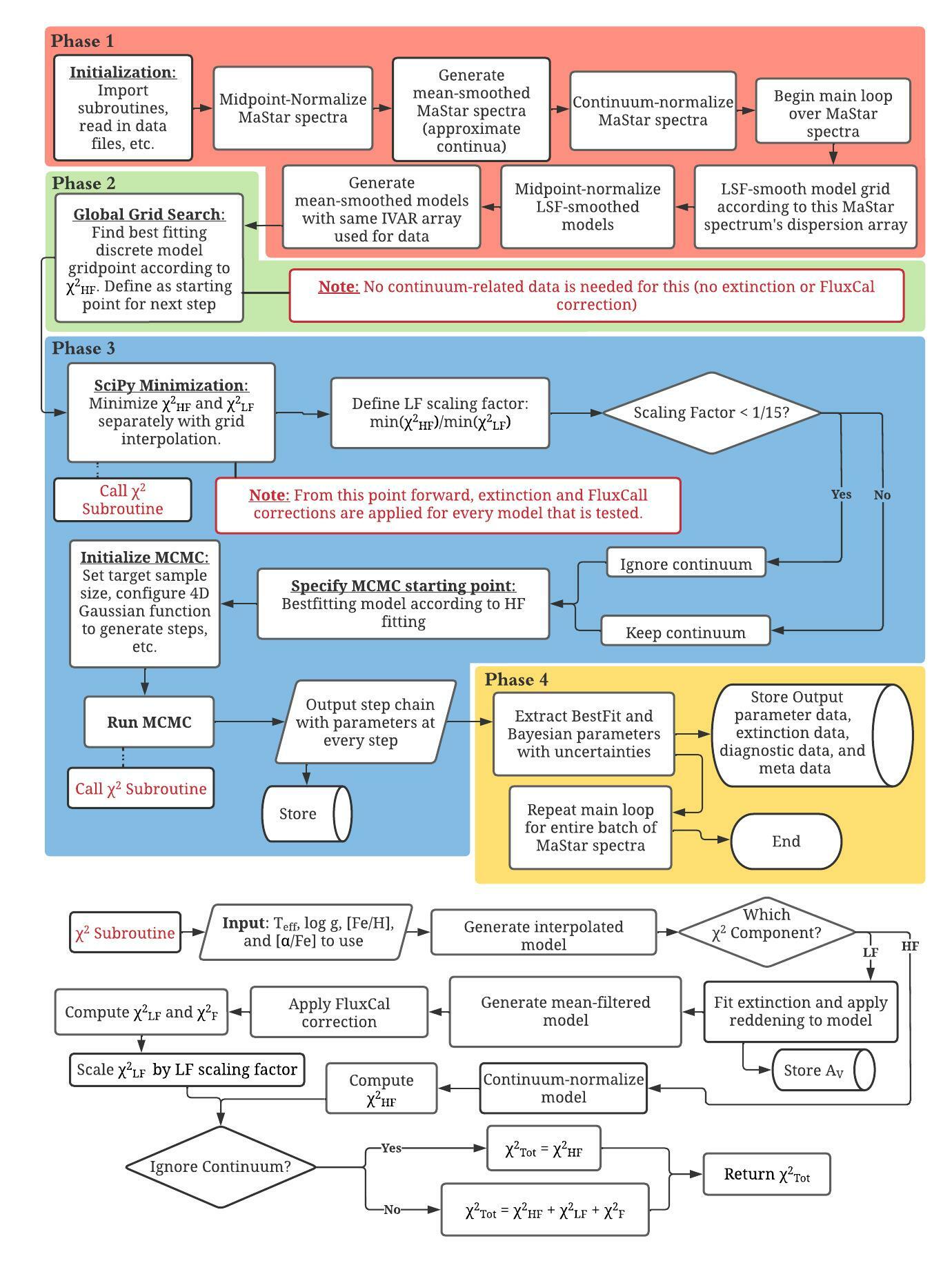}
        \caption{Flowchart detailing our parameter-determination procedure (upper) with an additional chart detailing the subroutine used for all $\chi^{2}$ calculations following the initial global grid search (lower).}
        \label{fig:flowchart_fig}
    \end{figure*}

   Figure \ref{fig:flowchart_fig} contains a flowchart describing these phases in further detail. From this, it is easy to see that the bulk of our method's nuance was applied within Phase 3. It was here that we considered several important but more subtle factors, such as $\chi^{2}_{\rm LF}$ scaling and rejection, and when extinction and flux-calibration correction were applied. We discuss these topics in detail below.
   
\subsection{Data}
\label{data}

    Data for MaStar was obtained at the Apache Point Observatory in New Mexico, alongside that of the aforementioned main surveys that make up SDSS-IV, using the Sloan 2.5-meter telescope \citep{Gunn06}. APOGEE-2, being a near-infrared H-band stellar survey, conducted observations during bright time. Since APOGEE-2 used its own infrared spectrograph, this presented an opportunity for MaStar to piggyback on APOGEE-2 observations using the unoccupied BOSS spectrograph \citep{Smee13} to observe at optical wavelengths simultaneously. This could be done using the MaNGA fiber bundles \citep{Drory15} to target stars. This setup allowed us to observe MaStar targets with a wavelength range of 3622-10,354\AA\ at a resolving power of  $R\sim1800$. To ensure high signal-to-noise ratios, MaStar targets were selected to be at least as bright as $17.5$ in either the g- or the i-band, and fainter than $8.1$ in both the g- and i-bands, according to preexisting data.
    
    As of its final data release, MaStar contains $25,683$ stellar spectra from $11,817$ unique science targets with a median signal-to-noise ratio of $110$. Each target received one or more observing visits, with a median of two visits and a maximum of 17 visits. Of these targets, $4,783$ received only one visit. In addition to the science targets, $33,583$ spectra from $12,477$ unique standard stars (targeted for flux calibration purposes) are also included in the final MaStar data file.
    
    A great deal of effort has been put into optimizing MaNGA and MaStar's flux calibration and LSF calibration. Flux calibration is considered to be accurate to within $3.9$\% in g-r, $2.7$\% in r-i, and $2.2$\% in i-z when compared to photometry \citep[Figure~5]{Yan_2019}, but is expected to vary slightly between spectra. The instrumental LSF, while having been acquired to similar accuracy \citep{law_2021}, does vary as a function of wavelength, and thus must be considered when conducting any comparison between MaStar spectra and high-resolution models.

\subsection{Model spectra}
\label{model_spectra}

    The model grid we employed was constructed using the ATLAS9-based model atmosphere database presented in \citet{meszaros2012} specifically for use with MaNGA and MaStar flux calibration, making it an excellent choice for our parameter-determination efforts. The line lists adopted for these models were the same as those described in \cite{BohlinM17}, though the grid is updated to have finer parameter sampling. The spectra cover a wavelength range of $3,000-11,000$ $\AA$ using vacuum wavelengths, with a resolution $R=10,000$. The spectra contain no rotational broadening, and the micro-turbulent velocity is the same as that used in the original model atmospheres, $2$ km/s.
    
\begin{figure}
        \centering
        \includegraphics[width=\columnwidth]
        {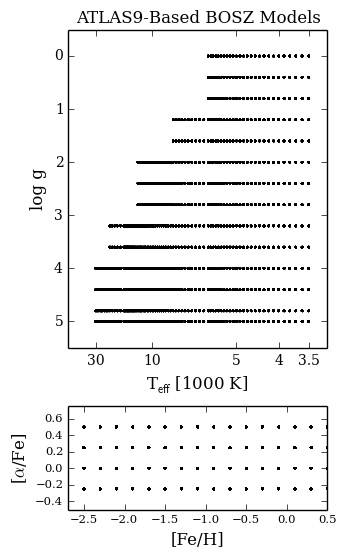}
        \caption{Atlas9-Based BOSZ model grid used for multicomponent $\chi^{2}$-fitting.}
        \label{fig:bosz_model_params}
    \end{figure}

    The grid covers most of MaStar's parameter footprint, ranging $3,500$ - $30,000$ K in temperature, $0$ - $5.0$ in $\log g$, $-2.5$ - $+0.5$ in [Fe/H], and $-0.25$ - $+0.5$ in [$\alpha$/Fe]. The grid we chose is moderately dense, with grid points having a spacing of $0.4$ dex in $\log g$ (with the exception of the two highest bins, $4.8$ and $5.0$), $0.2$ dex in metallicity, and $0.25$ dex in $\alpha$ abundance. The temperature grid spacing, however, is nonuniform, giving denser coverage at lower temperatures ($100$ K spacing below $6,000$ K), and becoming more sparse at higher temperatures ($200$ K spacing between $6,000$ K and $15,000$ K, and $1,000$ K spacing above $15,000$ K). For computational convenience, we found it necessary to modify the step sizes above $10,000$ K and in the $\log g$ dimension, making our final grid slightly less dense than the original that was provided to us. One could use a denser grid if desired, though it would be more computationally expensive. The BOSZ model atmospheres are also capable of representing multiple carbon abundances, but keeping [C/M]$=0$ was sufficient for our purposes. The layout of our final model grid is described by Table \ref{table: grid_table} and Figure \ref{fig:bosz_model_params}.
    
    \begin{table*}
    \caption{Atmospheric parameters of ATLAS9-based BOSZ model spectra} 
    \label{table: grid_table}                   
    \centering                                      
    \begin{tabular}{c c c c c c c c}          
    \hline\hline                        
    Parameter & Min & Max & Step & Parameter & Min & Max & Step\textsuperscript{a} \\
    \noalign{\smallskip}
    \hline
    \noalign{\smallskip}
    \hline                                   
        $[$Fe/H$]$ & $-$2.5 & 0.5 & 0.2 & & & &\\      
        $[\alpha$/M$]$ & $-$0.25 & 0.5 & 0.25 & & & & \\
        T$_{\rm eff}$ & 3500 & 6000 & 100 & $\log g$ & 0.0 & 5.0 & 0.4 \\
        T$_{\rm eff}$ & 6200 & 8000 & 200 & $\log g$ & 1.2 & 5.0 & 0.4 \\
        T$_{\rm eff}$ & 8200 & 12000 & 200 & $\log g$ & 2.0 & 5.0 & 0.4 \\
        T$_{\rm eff}$ & 12200 & 15000 & 200 & $\log g$ & 3.2 & 5.0 & 0.4 \\
        T$_{\rm eff}$ & 16000 & 20000 & 1000 & $\log g$ & 3.2 & 5.0 & 0.4 \\
        T$_{\rm eff}$ & 21000 & 30000 & 1000 & $\log g$ & 4.0 & 5.0 & 0.4 \\
    \hline    
    \multicolumn{5}{l}{\textsuperscript{a}\footnotesize{$\log g$ bins $0.0$ - $4.8$ have uniform step size of $0.4$, and the highest bin is $5.0$.}} \\
    \end{tabular}
    \end{table*}
    
\subsection{LSF model correction}
\label{lsf_model_correction}

    Our chosen set of model spectra were provided at a much higher resolution than that of the observed MaStar spectra ($R$ $\approx$ $1,800$), thus, in order to make an accurate pixel-by-pixel comparison between the observed MaStar spectra and the model spectra, it was necessary to both 1) convolve the model spectra with MaStar's instrumental broadening kernel to bring them to the correct resolution, and 2) resample the model spectra according to MaStar's wavelength array. Neglecting this first step would have led to disagreement in line profiles and different degrees of blending of unresolved lines.
    
    Since the LSF varies with wavelength in a way that is unique to each visit, LSF convolution had to be done on a pixel-by-pixel and spectrum-by-spectrum basis. To accomplish this, we constructed a 3D array containing the full set of model spectra convolved with Gaussian kernels with 17 successive $\sigma$ values. We then interpolated this array using the pixel-by-pixel dispersion provided with each MaStar spectrum in order to obtain a version of the model grid that was smoothed and resampled to a resolution that matches the data. We could then store this array and call it for each new MaStar spectrum to provide a model grid that had been convolved appropriately to match that particular spectrum's wavelength-dependant LSF.
    
    For each MaStar spectrum, the appropriate LSF-smoothed set of model spectra was generated up front during Phase 1, as represented in the flowchart in Figure \ref{fig:flowchart_fig}. We then used this smoothed model set in subsequent steps to generate new model spectra via interpolation.

\subsection{Continuum normalization}
\label{continuum_normalization}

    Characterizing the large-scale shape of stellar spectra is a necessary step for most traditional spectroscopy-based parameter-determination methods, which typically perform some sort of continuum normalization on observed spectra prior to fitting. The reason for doing this is that imperfect flux calibration can have an impact on the overall spectral shape, and removing the low-frequency spectral features that are susceptible to such problems helps to mitigate this. However, the broadband, low-frequency shape of a spectrum contains information that can be used for constraining atmospheric parameters. One of our goals was to make the best possible use of this information, which many previous efforts have disregarded entirely. Throughout this paper, the term ``continuum'' is used to refer to the broadband shape of a stellar spectrum, as opposed to high-frequency features such as atomic and molecular lines. This is not necessarily the same as the physical continuum that one would observe in a spectrum with very high resolution.
    
    To capture the broadband shape, we used a simple mean-filter smoothing routine that was weighted by the pixel-by-pixel inverse variance provided with the MaStar spectra. This provided an easy way of smoothing an observed spectrum down to an approximation of its broadband continuum shape, preserving its large-wavelength-scale features. We preferred this method over the use of b-spline or polynomial functions, since it was computationally more efficient than b-spline fits and more accurate than polynomial fits. In addition, since the smoothing was weighted by the pixel-by-pixel uncertainties unique to each observed spectrum, we could perform an identical smoothing on all the model spectra using the same inverse variance vector as the weight function, resulting in an unbiased comparison. This provided an advantage over many other continuum-determination methods.
    
    This inverse variance weighted mean smoothing routine was essentially a boxcar running mean calculation that uses a window size of 300 pixels (260-690 $\r{A}$, depending on the wavelength regime), represented by RM in Equation \eqref{myeqn1}:
    
    \begin{equation}
    \tilde{F}(\lambda) = \frac{RM(F_{\lambda}/\sigma^2)}{RM(1/\sigma^2)}, \label{myeqn1}
    \end{equation}
    
    \noindent{where $\tilde{F}(\lambda$) is the mean-smoothed flux, and $\sigma^2$ is the unique pixel-by-pixel dispersion provided for each MaStar spectrum. An analogous running sum was also performed on the inverse variance associated with each spectrum, giving the appropriate smoothed inverse variance at each pixel, which was needed for the subsequent $\chi^{2}$ fitting.}

    \begin{figure}
        \centering
        \includegraphics[width=8.2cm]{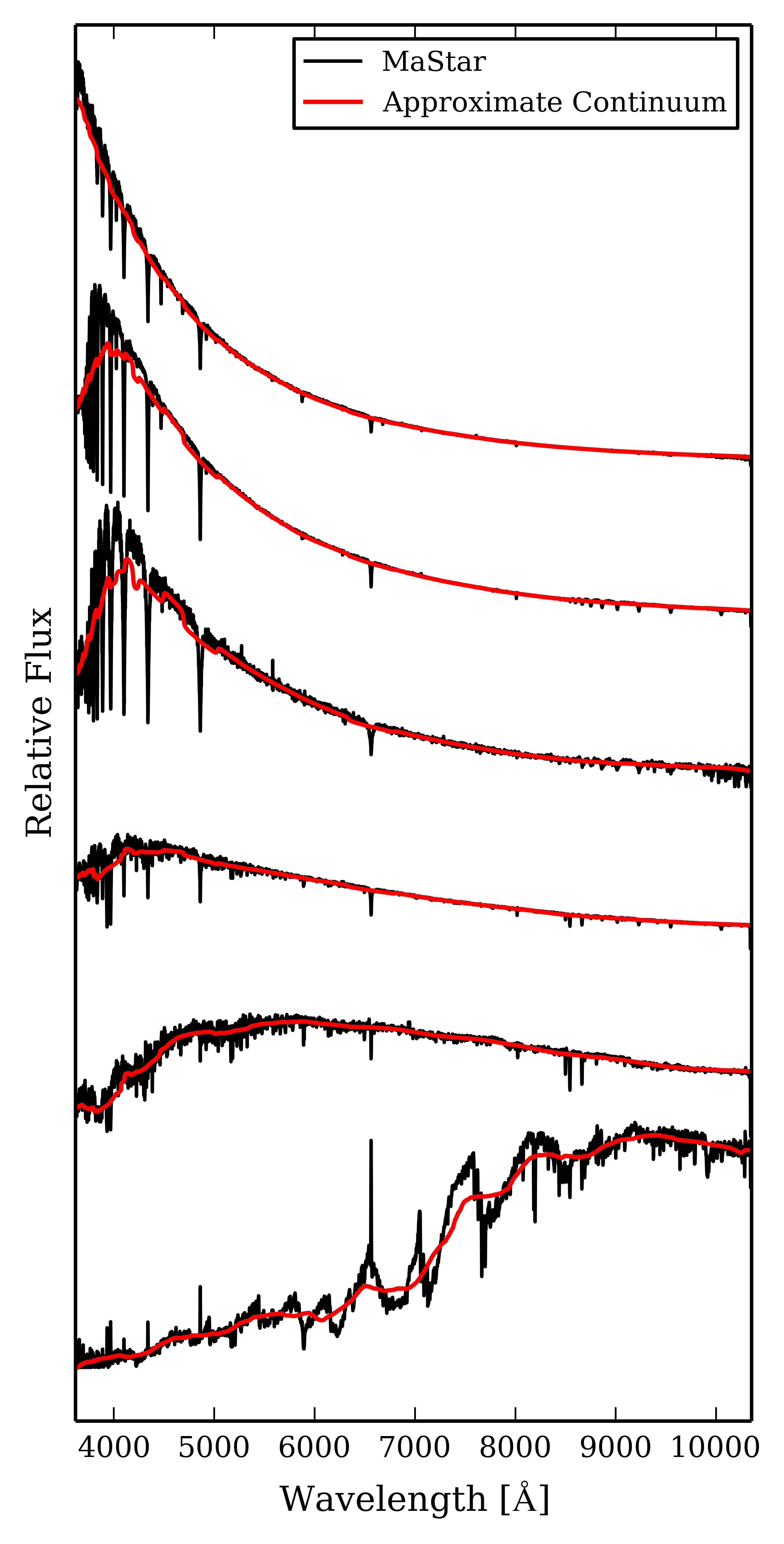}
        \caption{Several examples of observed stellar spectra (black) with their inverse-variance-weighted mean smoothed counterparts (red). The latter was used for continuum normalization and for direct comparison with model continua. Temperatures range from hottest to coolest from top to bottom.}
        \label{fig:continuum_approx_fig}
    \end{figure}

    The window size essentially set an upper limit on the width of features that could remain in the normalized spectra. As such, the width of 300 pixels was chosen in an ad hoc manner in order to smooth out low-frequency features while leaving high-frequency features intact in the final normalized spectra. Future work may be needed to determine whether or not 300 pixels is truly the optimal width for this purpose, and whether the optimal width ought to vary significantly between different spectral types. However, we found this choice to be perfectly acceptable for our purposes. For hotter stars ($T_{\rm eff} \geq 6,000$ K), this method typically gave a very accurate representation of the continuum. It also remained effective at lower temperatures, where molecular features become prominent, but the smoothed spectra in such regimes became less representative of the true physical continua due to the very wide molecular features. However, this did not impact the ability of the routine to find a reasonable match between the data and the models, since their continua were treated identically.
    
    Figure \ref{fig:continuum_approx_fig} shows several examples of how the continuum representation behaves for a variety of spectral types. It is also worth noting that higher-frequency spectral features, such as prominent Hydrogen absorption lines, metal lines, and line blanketing could often cause small wiggles in the continuum. This can be seen in some of the intermediate-temperature spectra shown in Figure \ref{fig:continuum_approx_fig}. Since the data and the models were treated in exactly the same way, these features did not bias the fitting at all.
    
    Once we obtained the continuum, we could then generate a continuum-normalized spectrum by dividing the original spectrum by the continuum. This normalized spectrum could then be used for high-frequency spectral fitting, and the continuum could be used for low-frequency spectral fitting. This continuum calculation and normalization for the data and models were done in Phase 1, as shown in Figure \ref{fig:flowchart_fig}. 

\subsection{Defining the primary components of $\chi^{2}$}
\label{primary_components_of_chisq}

    The agreement between a given MaStar spectrum and a model spectrum was characterized by $\chi^{2}_{\rm Total}$, which we defined as the sum of two main constituents, $\chi^{2}_{\rm HF}$ and $\chi^{2}_{\rm LF}$. These two components were obtained using simple reduced-$\chi^{2}$ calculations, measuring the agreement between the given spectra on a pixel-by-pixel basis, as shown in Equations \eqref{myeqn2} and \eqref{myeqn3}:
    
    \begin{equation}
    \chi^{2}_{\rm HF} = \frac{1}{N-1}\sum_{i}^{}\frac{(f_{i}-f_{\rm m,i})^{2}}{\sigma^{2}_{i}} \label{myeqn2}
    \end{equation}

    \begin{equation}
    \chi^{2}_{\rm LF} = \frac{1}{N-1}\sum_{i}^{}\frac{(C_{i}-C_{\rm m,i})^{2}}{\sigma^{2}_{C,i}} \label{myeqn3},
    \end{equation}
    
    \noindent{where N is the number of pixels in the wavelength array, $f_{i}$ and $f_{\rm  m,i}$ are the continuum-normalized flux for the data and the model, respectively, and  $C_{i}$ and $C_{\rm m,i}$ are the data and model continuum, respectively. $\sigma^{2}_{i}$ is the variance at pixel $i$ provided with the data, and $\sigma^{2}_{C,i}$ is the variance associated with the continuum, obtained via mean-smoothing, as discussed in Section \ref{continuum_normalization}.
    
    Despite the similarity in their construction, the values of $\chi^{2}_{\rm HF}$ and $\chi^{2}_{\rm LF}$ could differ greatly. It was typical for $\chi^{2}_{\rm LF}$ to be significantly greater than $\chi^{2}_{\rm HF}$, often exceeding it by an order of magnitude or more. This put $\chi^{2}_{\rm LF}$ in danger of dominating $\chi^{2}_{\rm HF}$, often leading to inaccurate estimates. This discrepancy in scale indicated that our formulation for $\chi^{2}_{\rm LF}$ as described by Equation \ref{myeqn3} does not properly take into account the covariance introduced by the mean-smoothing routine used to obtain $C_{i}$. This would typically result in $\sigma^{2}_{C,i}$ being significantly underestimated. We addressed this by introducing an ad hoc fix in the form of a corrective scaling factor, as shown in Equation \eqref{myeqn4}:
    
    \begin{equation}
    \bar{\chi}^{2}_{\rm LF} = \chi^{2}_{\rm LF}\frac{min(\chi^{2}_{\rm HF})}{min(\chi^{2}_{\rm LF})} \label{myeqn4},
    \end{equation}
    
    \noindent{where $\bar{\chi}^{2}_{\rm LF}$ refers to the scaled $\chi^{2}$ measurement associated with the low-frequency component of a particular model's spectral shape. This prevented the low-frequency fitting from dominating the high-frequency fitting, thus preventing our result from being driven to an incorrect part of parameter space.}
    
    To obtain a value for $min(\chi^{2}_{\rm HF})$ and $min(\chi^{2}_{\rm LF})$ for use in Equation \ref{myeqn4}, we used an optimization function \texttt{scipy.optimize} \citep{2020SciPy-NMeth}. This package was only used for calculating the scaling factor, not for finding the best fit with the combined $\chi^{2}$. We chose to use a different algorithm for minimizing $\chi^{2}_{\rm Total}$, since the \texttt{scipy.optimize} algorithm had a tendency to become trapped in local minima, which would ultimately lead to gridding artifacts in the final distribution.
    
    In a number of cases, we found that the mitigating scaling factor alone was not sufficient for addressing the problem of $\chi^{2}_{\rm Total}$ being dominated by $\chi^{2}_{\rm LF}$. This would occur predominantly for the cool main sequence ($T_{\rm eff}$ $<$ $4,000$ K) and at the tip of the red giant branch ($log$ $g$ $<$ $1$). We believe that the cause for this was a mismatch between the data and the models in the molecular features present in cool stars. As a result, $min(\chi^{2}_{\rm LF})$ could exceed $min(\chi^{2}_{\rm HF})$ by a factor of 20 or more, leading to a tendency for the algorithm to underestimate $\log g$ in cool dwarfs. To mitigate this, we imposed a minimum permissible scaling factor of $1/15$. That is, if $min(\chi^{2}_{\rm LF})$ exceeded $min(\chi^{2}_{\rm HF})$ by more than a factor of 15, then we chose to disregard all continuum-related components of $\chi^{2}_{\rm Total}$ entirely, including both $\chi^{2}_{\rm LF}$ and $\chi^{2}_{\rm F}$. In these cases, the only point where the continuum was still used was in the continuum-normalization step. It is important to note here that this effect was not due to a shortcoming of the fitting technique. Rather, it suggests some overall disagreement between the data and the models in this parameter regime. We discuss this in more detail in Section \ref{wavelength_regime_weighting}.
    
    These $\chi^2$ components were computed at several different stages throughout the routine. In the global grid search performed in Phase 2, only $\chi^{2}_{\rm HF}$ was used, since the goal of this step was only to generate an initial estimate of the stellar parameters to later be used as a starting point for the MCMC. Throughout Phase 3, all $\chi^2$ calculations were handled by a subroutine described by the lower section of the flowchart in Figure \ref{fig:flowchart_fig}.

\subsection{Extinction fitting}
\label{extinction_fitting}

    Accounting for the reddening effect of interstellar extinction was critical for our approach, since it can dramatically affect the broadband shape of a spectrum. Hence, when attempting to fit a given MaStar spectrum, we had to consider the possibility that its shape may have been significantly altered by the presence of interstellar dust along the line of sight.
    
    One possible approach to accounting for extinction involves the use of 3D dust maps (such as those presented in \citealt{Green_2019}) to estimate the line-of-sight extinction. This usually gives a reliable extinction estimate, provided the distance and the distribution of dust in the vicinity of the star are sufficiently well known. The number of MaStar targets that satisfy this condition is quite large, thanks to projects such as the Gaia mission \citep{gaia_2018}, which can provide accurate distance estimates using parallax measurements, especially for relatively nearby stars \citep{Bailer_Jones_2018}. However, this dust map approach becomes less reliable for distant stars and hot stars. Distant stars are problematic because distance estimates become more uncertain as the fractional uncertainties in their parallax measurements become larger. Hot stars are also problematic because they tend to be located within star forming regions, where the extinction level jumps significantly. As a result, a small error in distance could cause a large error in extinction. Given that upper main-sequence stars and giant stars make up an important portion of MaStar's target set, we chose not to rely on 3D dust map information for extinction fitting.
    
    Rather than trusting literature values for the extinction along a given line of sight, we instead incorporated extinction fitting into the model search by computing the extinction that a given model being tested would need in order to match the data using extinction curves provided in the literature. To illustrate this, consider the usual expression used to compute the extincted flux, $F_{\rm E}$, given an extinction value, $A_{\rm V}$:

    \begin{equation}
    F_{\rm E, \lambda} = QF_{\lambda}10^{-0.4\frac{A_{\lambda}}{A_{\rm V}}A_{\rm V}} \label{myeqn5}.
    \end{equation}
    
    \noindent{In principle, there is a proportional constant that depends on the normalization differences between the data and the model, which we represent with $Q$. The fraction $A_{\lambda}/A_{\rm V}$ is provided by the Fitzpatrick extinction curves \citep{Fitzpatrick_2019}. By rearranging Equation \ref{myeqn5}, we can obtain an expression that is linear in $A_{\rm V}$:}
    
    \begin{equation}
    \log (F_{\rm E, \lambda}/F_{\lambda}) = {-0.4\frac{A_{\lambda}}{A_{\rm V}}A_{\rm V}} + \widetilde{Q} \label{myeqn6}.
    \end{equation}

    \noindent{Substituting $F_{\rm E,\lambda}$ with the MaStar spectrum for which we wish to obtain an extinction estimate ($F_{\rm MaStar}$) and $F_{\lambda}$ for a given model spectrum that we wish to compare ($F_{\rm Model}$), and plotting $log_{10}(F_{\rm Model}/F_{\rm MaStar})$ as a function of $0.4\frac{A_{\lambda}}{A_{\rm V}}$, typically yields a near straight-line curve that can be fit with a linear function having a slope that can be used as an $A_{\rm V}$ estimate, as shown in Figure \ref{fig:extinction_fig}. This function also has a vertical offset $\tilde{Q}$, which we can discard. In developing this method, we took special care here to limit the range of the fitting to a wavelength range that was less sensitive to model mismatch, had high signal-to-noise ratios, and was free of potential data issues. For this reason, we only performed the fit over the wavelength range $4,775-9,500$ $\r{A}$.}
    
    It is important to note that this technique faced a disadvantage at low temperatures. For cool dwarf and cool giant stars, any deviation seen in $log_{10}(F_{\rm Model}/F_{\rm MaStar})$ from a linear function was often dominated by deviations due to significant model mismatches, which occurred frequently for cool stars, since the models provided poor fits to the large molecular bands in cool stellar spectra. Because of this, we recommend that our extinction estimates be used with caution in certain parameter regimes. Such problematic cases can usually be identified by their high error estimates. We discuss this in further detail in Section \ref{extinction_comparison}. We performed our extinction-fitting procedure every time $\chi^{2}_{\rm LF}$ was calculated throughout Phase 3 of the routine, as described in Figure \ref{fig:flowchart_fig}. It was built into the $\chi^{2}$ subroutine.
    
    \begin{figure}
        \centering
        \includegraphics[width=\columnwidth]
        {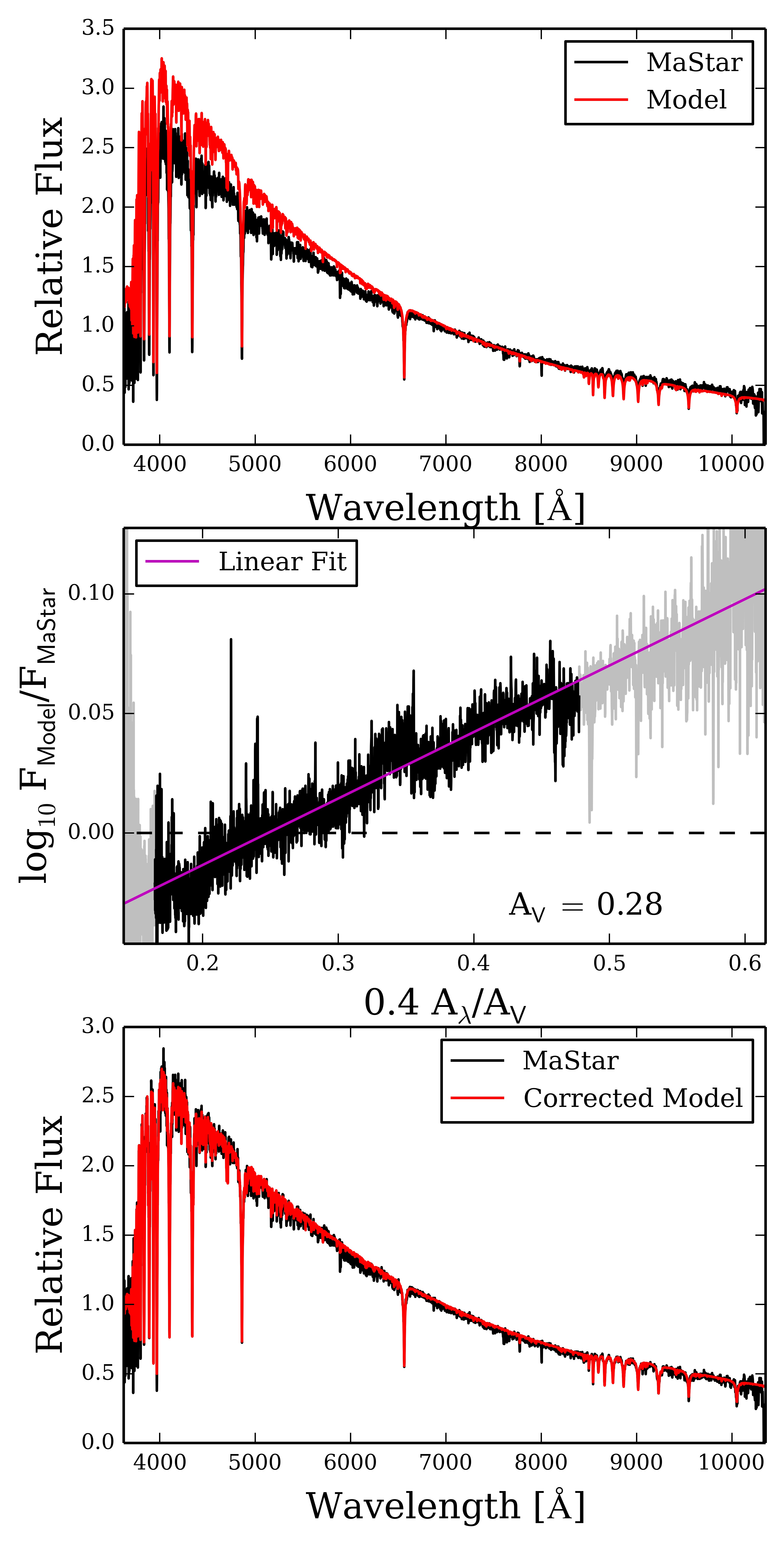}
        \caption{Extinction fitting using a linear fit of the $log_{10}$ of the flux ratio of a given data-model pair. The fitting was limited to a wavelength range that was relatively insensitive to model mismatch, had good S/N, and was free of data issues. $R$ $=$ $3.1$ was assumed.}
        \label{fig:extinction_fig}
    \end{figure}

\subsection{Flux-calibration systematics}
\label{flux_calibration_systematics}

    While MaStar's flux calibration is considered to be very accurate, even slight systematics can have a significant impact on low-frequency spectral features. In practice, this means that, at any given wavelength, an observed spectrum's continuum shape may be off by as much as $5-15$\% \citep[Figure~6]{Yan_2019}, so we had to adjust our low-frequency fitting accordingly.
    
    In order to allow for this possible deviation from a star's true spectral shape for each data-model pair of spectra, we evaluated the extent to which the flux calibration would have to be off in order to account for any disagreement. This could be characterized by the ratio between the data and model continua of each spectrum, $C_{MaStar}/C_{BOSZ}$, as a function of wavelength. In this representation of the flux-calibration residuals, a ratio that is equal to 1 at every wavelength would correspond to a perfect broadband fit, and no inaccuracy in flux calibration would be suggested. However, in a realistic case, we would expect that this ratio could deviate between $\sim 0.85$ and $\sim1.15$ at either end of the wavelength range, while still corresponding to an acceptable fit. For each pair of data and model spectra, we fitted a fifth-order polynomial to the continuum ratio to approximate the flux-calibration residual vector. The model spectrum was then divided by this fifth-order polynomial to make it conform to the data, in effect removing the flux calibration residuals present in the fit, and allowing for a much more fair comparison between the data and model continua.
    
    The obvious danger involved with this procedure was the possibility that we could be forcing poorly fitting models to fit the data. For this reason, we chose to impose some limitation on what level of disagreement we considered to be explainable by flux-calibration systematics. For example, if a data-model pair had a continuum ratio that extended much higher than 1.15 or much lower than 0.85, we needed to have a way of ruling out that model as a reasonable choice. To do this, we characterized the extremeness of the fifth-order polynomial correction with its own component of $\chi^{2}_{\rm Total}$, which we refer to as $\chi^{2}_{\rm F}$. This component has the form:
    
    \begin{equation}
    \chi^{2}_{F} = \frac{\ln(P_{Blue}(max(|P_{\rm Blue} - 1|))^{2} +       \ln(P_{Red}(max(|P_{\rm Red} - 1|))^{2}}{2\ln(1.21)^{2}} \label{myeqn7},
    \end{equation}
    
    \noindent{where $P_{\rm Blue}$ and $P_{\rm Red}$ are the fifth-order polynomial correction being applied to the model, evaluated for $\lambda$ < 6980 $\r{A}$ and $\lambda$ > 6980 $\r{A}$, respectively. Each section is evaluated at its point of maximum absolute deviation from 1 at the blue and red ends of the wavelength range. This would typically amount to a very small correction to $\chi^{2}_{\rm Total}$ near the minimum of the distribution, but was used to ensure that models that would require an unreasonable flux-calibration correction were deprioritized. This flux-calibration correction procedure was performed every time $\chi^{2}_{\rm LF}$ was calculated in Phase 3, as described in Figure \ref{fig:flowchart_fig}. It was built into the $\chi^{2}$ subroutine that is described in the lower part of the figure.}

\subsection{Markov chain monte carlo (MCMC) algorithm}
\label{mcmc_algorithm}

    With all of the components of $\chi^2_{\rm Total}$ defined, the next task was to search the 4D parameter space for the model that gives the minimum $\chi^2_{\rm Total}$. Our approach to this used a simple Markov Chain Monte Carlo (MCMC) algorithm to sample the $\chi^2_{\rm Total}$ distribution. The MCMC algorithm is a well known tool for sampling multidimensional parameter spaces, and has the advantage of being easy to understand and operate.
    
    The MCMC algorithm essentially amounts to a random walk, in which a starting point is chosen, and each new step is proposed by drawing randomly from a given distribution and accepted with a $\chi^{2}$-dependent probability. We chose our walker's starting point to be the location of the model grid point that provided the minimum $\chi^{2}_{\rm HF}$, which was obtained during the global grid search step in Phase 2 of the procedure. This ensured that the starting point for the walker was usually not far from the true $\chi^{2}$ minimum that we wished to sample. Each subsequent proposed step was selected from a 4D Gaussian distribution with dimensional standard deviations initially determined by the model grid spacing. The proposed step was accepted into the step chain with a probability given by $exp(\chi^2_{\rm old}-\chi^2_{\rm new})$ or 1 (whichever was smaller). This made the walker naturally tend toward better-fitting models without being limited to stay within local minima. We continuously adjusted the walker's step size to maintain an acceptance rate close to 0.234, which is considered to be optimal for n-dimensional MCMC algorithms \citep{Gelman1997}. 
    
    The walker was typically required to take several hundred initial steps to ensure that it had successfully located and begun sampling the vicinity of the absolute minimum. This initial series of steps (often called the burn-in period) had to be discarded before the step chain could be treated as a proper sampling of the $\chi^{2}$ distribution. For our purposes, a chain of 2000 accepted steps following the burn-in period was considered to be a sufficient sample, at which point the algorithm would terminate. To improve efficiency, we also allowed the MCMC to stop early if it had found enough data points close to the absolute minimum of the distribution. If the step chain accumulated 1000 accepted steps with $\chi^{2}$ values no greater than that of the running minimum $\chi^{2}$ + 1, we permitted the MCMC to stop before the chain reached 2000 accepted steps. Prioritizing this subset of points near the minimum of the distribution also proved helpful in extracting parameter and error estimates at a later point, as discussed in Section \ref{extracting_parameters_and_uncertainties}. We specified a third condition to stop the MCMC if it had been compiling a single chain for more than 20 minutes. The MCMC occasionally had difficulty finding a well-defined minimum in the $\chi^{2}$ distribution, and the problem could usually be traced back to some artifact in the data. This was extremely rare, and is only known to have affected two of the $59,266$ spectra. Any such problematic cases were flagged appropriately in the final parameter set. For several illustrative examples of the MCMC's behavior and convergence on a set of minimum-$\chi^{2}$ parameters, we refer the reader to Appendix \ref{mcmc_behavior}. The MCMC was run in Phase 3. At each step, it called the $\chi^{2}$ subroutine to compute $\chi^{2}_{Tot}$ for the newly generated model spectrum.
    
\subsection{Wavelength regime weighting} 
\label{wavelength_regime_weighting}

    Through inspection of the fitting performance, we found that the BOSZ model grid faced difficulties fitting intermediate-scale spectral features for cool stars with $T_{\rm eff} < 4,000$ K (e.g., TiO molecular bands). The scale of these features was problematic, since it was large enough to affect the broadband $\chi^{2}$ fitting as well as the narrow-band component. The former was partially addressed by the continuum rejection described in Section \ref{primary_components_of_chisq}, in which $\chi^{2}_{\rm LF}$ and $\chi^{2}_{\rm F}$ were excluded from $\chi^{2}_{\rm Total}$ when certain conditions were met. However, the direct impact on $\chi^{2}_{\rm HF}$ proved to be sufficiently strong as to still affect final parameter estimations for cool stars. A simple solution to this problem would be to mask the most problematic features and exclude them from the $\chi^{2}_{\rm HF}$ fitting entirely. However, this would run a major risk of discarding useful data in regions of parameter space where molecular bands are not a problem. 
    
    The approach we took was to de-weight pixels below $7,800$ $\r{A}$ universally, forcing them to have a smaller contribution to $\chi^{2}_{\rm HF}$ than those at the red end of the spectra by a factor of $C$, as shown in Equation \eqref{myeqn10}. This can be viewed as a further decomposition of $\chi^{2}_{\rm HF}$ into blue and red parts:
    
    \begin{equation}
    \chi^{2}_{\rm HF} = C\chi^{2}_{\rm Blue} + \chi^{2}_{\rm Red}. \label{myeqn10}
    \end{equation}

    This was an ad hoc approach that we chose to take based on trial and error, allowing the final distribution and its agreement with external calibration sets to determine which value of $C$ we adopted. Based on a series of tests, we found $C=0.1$ to produce satisfactory results in low-temperature regimes, such as the cool main sequence and the tip of the red giant branch. Our reason for choosing $7,800$ $\r{A}$ as the wavelength cutoff for this weighting scheme is the presence of large molecular bands that are common in cool dwarfs below this limit. An example case exhibiting some of these features is shown in Figure \ref{fig:TiO_example}.
    
    The effect this had on the parameter distribution in the cool main sequence and the red giant branch can be seen in Figure \ref{fig:unweighted_v_weighted_fig}, which compares results obtained with and without the wavelength weighting. For the cool main sequence, $\log g$ underestimation was greatly reduced by the introduction of the weighting. Other changes include the split in the upper red giant branch near $T_{\rm eff}$ = $4,500$ K being filled in, and the line of metal-rich giants continuing to the tip of the red giant branch, as opposed to stopping abruptly at $log$ $g\sim1.8$. Further effects from this are discussed in Section \ref{external_comparison}. This weighting procedure was incorporated in every $\chi^{2}_{\rm HF}$ calculation throughout the entirety of the parameter-determination process.
    
    \begin{figure*}
        \centering
        \includegraphics[width=\textwidth]
        {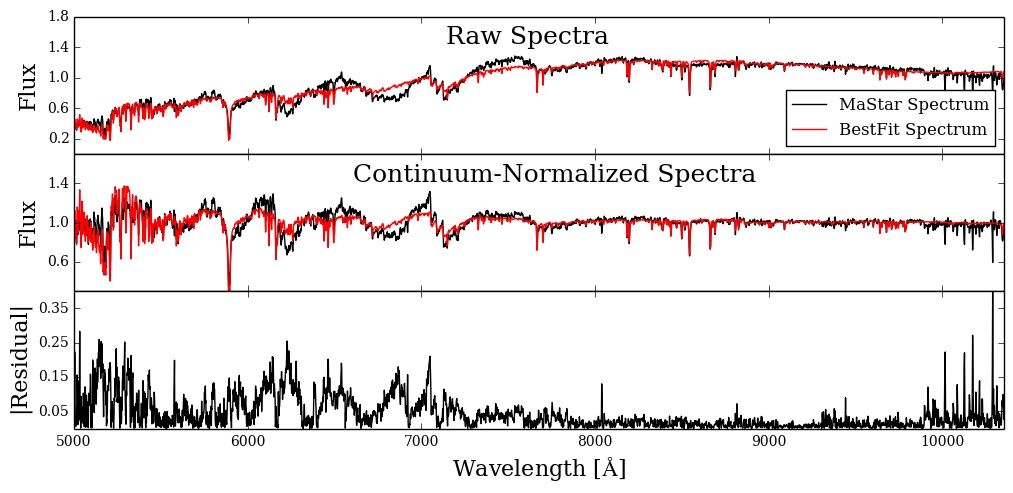}
        \caption{Example of a cool dwarf spectrum with prominent TiO bands below $7,800$ $\r{A}$, which our model set struggled to represent accurately. The raw spectrum, along with its corresponding BestFit model, is displayed in the top panel. The continuum-normalized spectrum and model are displayed in the middle panel. The absolute residuals from this fit are shown in the bottom panel, and show significantly better agreement in the $7,800$ $-$ $9,900$ $\r{A}$ range.}
        \label{fig:TiO_example}
    \end{figure*}
    
    \begin{figure}
        \centering
        \includegraphics[width=\columnwidth]
        {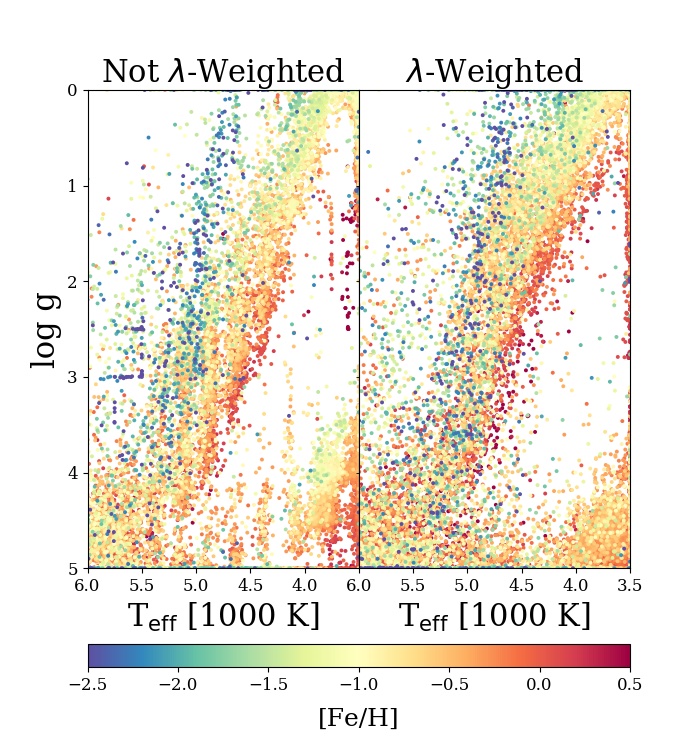}
        \caption{Comparison between MaStar's parameter distribution before (left) and after (right) introducing wavelength weighting to give higher priority to redder wavelengths. Deprioritizing fitting spectral features below $7,800$ $\AA$ led to better performance at lower temperatures.}
        \label{fig:unweighted_v_weighted_fig}
    \end{figure}

\subsection{Extracting parameters and uncertainties}
\label{extracting_parameters_and_uncertainties}

    For each MaStar spectrum that was processed by our parameter pipeline, two separate methods were used to extract the desired atmospheric parameters, $T_{\rm eff}$, $\log g$, [Fe/H], and [$\alpha$/Fe], from the data collected by the MCMC. The simpler method involved adopting the parameters associated with the best-fitting model encountered by the MCMC (that is, the one with the minimum $\chi^{2}_{\rm Total}$), and computing errors based on the sample contained in the step chain. We refer to these parameters as the ``BestFit'' parameters. For an alternative method, we used the MCMC step chain to compute a set of likelihood-weighted mean parameters. Since this method closely resembles the standard Bayesian approach to parameter and error estimation using a set of synthetic spectra, we refer to these results as our ``Bayesian'' parameters. Since the latter approach was based on an ensemble of model fits, rather than a single minimum-$\chi^{2}$ model, it is important to note that the final parameter estimate would depend more strongly on the shape of the distribution sampled by the MCMC. This is important to remember for irregular cases, in which the MCMC sampled a section of parameter space that contained two or more local minima. The possibility of the MCMC step chain repeatedly running into a model grid boundary, yielding an artificially truncated sample, was also likely to bias the sample resulting in an unreliable estimate. Because of this, Bayesian parameters positioned near the grid boundaries should be treated with caution.
    
    Both approaches to parameter extraction make use of the mathematical likelihood, which could be calculated directly from the $\chi^{2}_{\rm Total}$ distribution:
    
    \begin{equation}
    L_{i} = e^{-\chi^{2}_{\rm Total,i}} \label{myeqn11},
    \end{equation}
    
    \noindent{where $L_{i}$ represents the likelihood of a single model $i$ being the correct representation of the MaStar spectrum in question. This acted as a statistical weight when computing the weighted mean of the distribution in each parameter dimension. As such, models that yielded high $\chi^{2}_{\rm Total}$ contributed very little to the final result, and the opposite was true for models that yielded low $\chi^{2}_{\rm Total}$. The likelihood was used to construct the first, second, and third order moments to be used in subsequent calculations:}
    
    \begin{equation}
    \begin{multlined}
    I_{0} =\sum_{i}L_{i},\ \ \ 
    I_{1} = \sum_{i}x_{i}L_{i},\ \ \ 
    I_{2} = \sum_{i}x_{i}^{2}L_{i} \label{myeqn12},
    \end{multlined}
    \end{equation}

    \noindent{where, above, we use the symbol $x$ to represent whichever of the four desired parameters is being evaluated.
    
    As described previously, the BestFit parameters were extracted by adopting the parameters associated with the lowest $\chi^{2}_{\rm Total}$ encountered by the MCMC step chain:
    
    \begin{equation}
    \begin{multlined}
    (T_{\rm eff})_{\rm BestFit} = T_{\rm eff}|_{\rm min(\chi^{2}_{\rm Total})} \\ \\
    (log\ g)_{\rm BestFit} = log\ g|_{\rm min(\chi^{2}_{\rm Total})} \\ \\
    ([Fe/H])_{\rm BestFit} = [Fe/H]|_{\rm min(\chi^{2}_{\rm Total})} \\ \\
    ([\alpha/Fe])_{\rm BestFit} = [\alpha/Fe]|_{\rm min(\chi^{2}_{\rm Total})}. \\ \\
    \label{myeqn13}
    \end{multlined}
    \end{equation}

    \noindent{Errors associated with the BestFit parameters were then calculated using the moments given in Equation \eqref{myeqn12}:
    
    \begin{equation}
    \begin{multlined}
    \sigma_{x,\rm  BestFit}^{2} = \frac{\sum_{i}(x_{i}-x_{\rm BestFit})^{2}L_{i}}{\sum_{i}L_{i}} \\
    = \frac{\sum_{i}x_{i}^{2}L_{i}}{\sum_{i}L_{i}} - 2x_{\rm BestFit}\frac{\sum_{i}x_{i}L_{i}}{\sum_{i}L_{i}} + x_{\rm BestFit}^{2} \\
    = \frac{I_{2}}{I_{0}} - 2x_{\rm BestFit}\frac{I_{1}}{I_{0}} + x^{2}_{\rm BestFit}. \label{myeqn14}
    \end{multlined}
    \end{equation}

    The Bayesian parameter and error estimates were also calculated using the moments given in Equation \eqref{myeqn12}:
    
    \begin{equation}
    \begin{multlined}
    x_{\rm Bayesian} = \left \langle x_{i} \right \rangle = \frac{I_{1}}{I_0} \\
    \sigma_{x,\rm Bayesian}^{2} = \left \langle x_{i}^{2} \right \rangle - \left \langle x_{i} \right \rangle^{2} \\
    = \frac{I_{2}}{I_0} - \left ( \frac{I_{1}}{I_0} \right )^{2}. \label{myeqn15}
    \end{multlined}
    \end{equation}

    \noindent{Limiting the set of points we use in the preceding calculations to those with $\chi^{2}_{\rm Total}$ $\leq$ $min(\chi^{2}_{\rm Total}) + 1$ helped to ensure that the results were not artificially influenced by outlying entries in the MCMC step chain, or in instances in which the step chain ran into the grid boundary, resulting in an asymmetric sample. The first and fourth cases discussed in Appendix \ref{mcmc_behavior} are examples in which this latter scenario was of concern. All of the MCMC parameter and error calculations were performed in Phase 4 of the procedure.}

\section{Results}
\label{results}

\subsection{Parameter distribution and uncertainty estimates}
\label{raw_parameter_distribution_and_uncertainty_estimates}

    Figures \ref{fig:bestfit_raw_params}-\ref{fig:alpha_raw_params} show the final 4D distributions of the BestFit and Bayesian parameters (excluding those for standard stellar targets) prior to imposing any quality control cuts. While the two distributions bear an overall resemblance, there are several distinctions.

    Looking first at the BestFit parameters, the $\log g$ versus $T_{\rm eff}$ distribution (analogous to a traditional Hertzsprung-Russell Diagram or Color-Magnitude Diagram) shows most of the features one would expect in MaStar's footprint: a distinct upper and lower main sequence, a red giant branch, and a horizontal branch. Several obvious artifacts are present. For example, our grid's lower temperature limit and upper surface gravity limit had an impact on the shape of the cool main sequence. This forced it to take an artificial downward turn in $\log g$ (upward on the plot) at very low temperatures. As discussed below, this problem would be much worse without our decision to exclude the continuum-based components of $\chi^{2}_{\rm Total}$ from the final fit in certain cases. Based on comparison with APOGEE parameters, we believe that some of these stars ought to extend to lower temperatures and higher surface gravity than are permitted by our model grid. This may be addressed in the future by supplementing our model set with cooler spectra, and perhaps allowing some limited extrapolation beyond the boundaries. The tip of the red giant branch was also affected by the model grid's limitations, producing the feature extending to $log$ $g\sim2.9$ near $T_{\rm eff}$ $=$ $3,500$ K. We believe that most of these targets belong at either slightly lower temperature, or slightly lower surface gravity than our model grid is able to accommodate.
    
    One other notable feature is the gap in the upper main sequence at approximately $9,250$ K. Based on external verification using Gaia color data, it is extremely unlikely that this feature is due to any target selection effect on the part of MaStar. As we discuss in Section  \ref{inaccuracies_in_the_model_spectra}, we believe that this is most likely due to inaccuracies in the BOSZ model spectra.
    
    The subsequent [Fe/H] versus $T_{\rm eff}$ and [$\alpha$/Fe] versus [Fe/H] distributions (the latter shown in Figure \ref{fig:alpha_raw_params}) also look reasonable. The $\alpha$ abundance distribution displays the expected bimodality described thoroughly in previous works relating to the ``chemical cartography'' of the Milky Way's disk, such as \cite{Hayden_2015}.

\begin{figure*}   
    \centering
    \includegraphics[width=\textwidth]
    {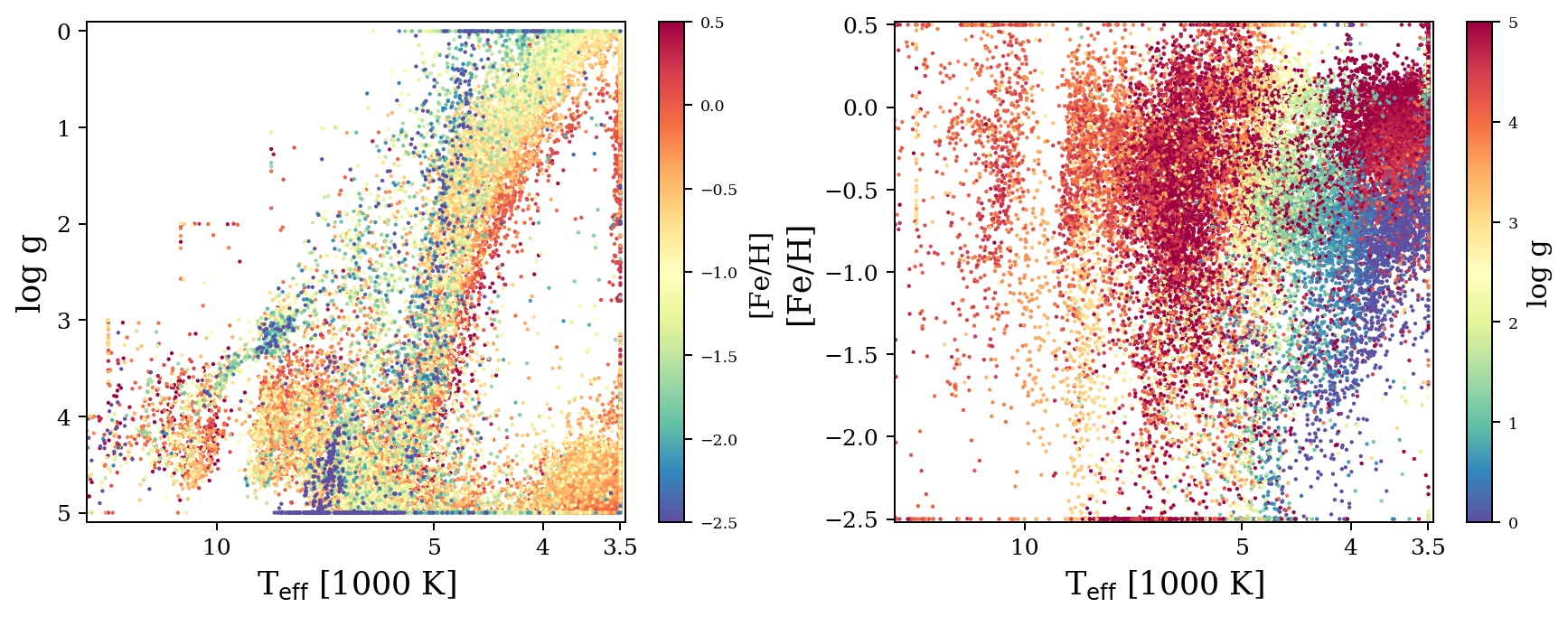}
    \caption{Distribution of MPL-11 BestFit parameters (standard stars excluded).}
    \label{fig:bestfit_raw_params}
\end{figure*}

    Looking next at the Bayesian parameter distributions, shown in Figure \ref{fig:bayesian_raw_params}, many of the features described previously are still present, including the $9,250$ K gap and cool main sequence turn, though the former appears less prominent. The reason for this is likely the aforementioned ``minimum-splitting'' effect that the Bayesian methodology was susceptible to. This refers to cases in which the final parameter estimate fell between two comparable local minima in the $\chi^2$ distribution. In such cases, specifically for the $9,250$ K gap, the fact that some points migrated inward to populate the gap is incidental, and does not necessarily indicate that a model generated at those parameters using our model grid would actually fit the data better. Also worth noting in this distribution, is the presence of far fewer cases of the parameters falling directly on the grid edge. This is an artificial effect due to cases in which the MCMC step chain ran into the grid boundary. We refer to this phenomenon as ``edge-clipping''. Since the Bayesian result was obtained from a weighted average of points surrounding the minimum of the distribution, this would often cause the final Bayesian parameter to be artificially pulled away from the grid boundary by the asymmetric distribution. Due to these problems with the Bayesian approach, we chose to adopt the BestFit results as our final parameters to recommend for external use. Figure \ref{fig:error_histograms_fig} shows the uncertainty distributions for the BestFit and Bayesian parameter estimates with median uncertainties included for the BestFit parameter set.

    \begin{figure*}
        \centering
        \includegraphics[width=\textwidth]
        {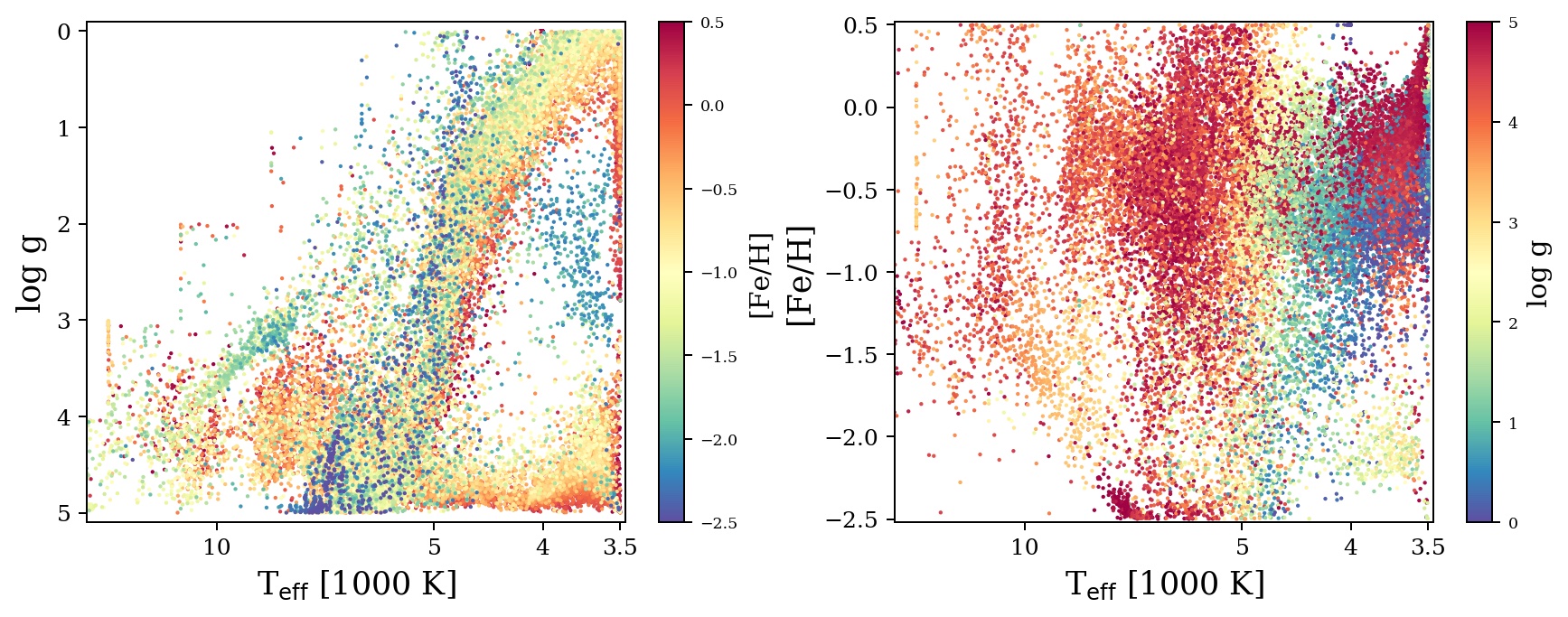}
        \caption{Distribution of MPL-11 Bayesian parameters (standard stars excluded).}
        \label{fig:bayesian_raw_params}
    \end{figure*}

    \begin{figure*}
        \centering
        \includegraphics[width=\textwidth]
        {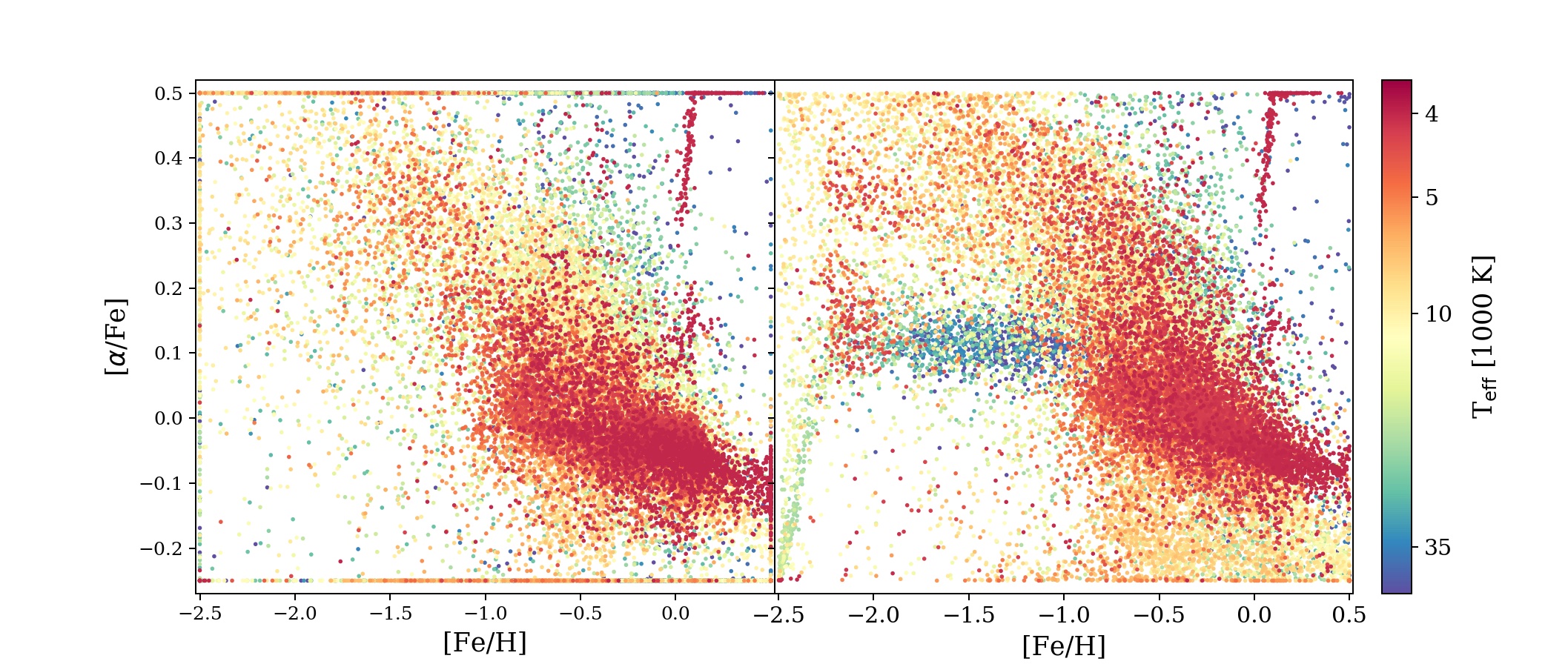}
        \caption{Distribution of MPL-11 BestFit (left) and Bayesian (right) parameters (Standard     Stars excluded).}
        \label{fig:alpha_raw_params}
    \end{figure*}

    \begin{figure*}
        \centering
        \includegraphics[width=\textwidth]
        {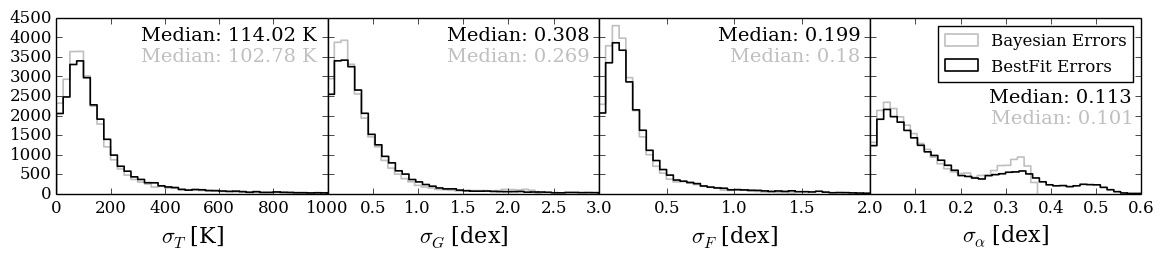}
        \caption{Histograms showing error distributions for BestFit and Bayesian parameter estimates for $T_{\rm eff}$, $\log g$, [Fe/H], and [$\alpha$/Fe] ($\sigma_{\rm T}$, $\sigma_{\rm G}$, $\sigma_{\rm F}$, and $\sigma_{\alpha}$ respectively). Standard targets have been excluded.}
        \label{fig:error_histograms_fig}
    \end{figure*}

\subsection{Internal consistency of error estimates}
\label{internal_consistency_of_error_estimates}

    Since many of MaStar's targets have multiple observations associated with them, it was relatively simple to evaluate the internal consistency of our error estimates. This allowed us to discern whether or not the uncertainties associated with the BestFit and Bayesian parameter sets were overestimated or underestimated, based on their consistency across multiple observations. Our preferred metric for evaluating this was the pairwise difference between the parameter estimates obtained from different observations, scaled by the quadratic sum of the errors associated with those estimates. We refer to this quantity as $\delta_{x_{ij}}$.
    
    \begin{equation}
    \delta_{x_{ij}} =  \frac{x_{i} - x_{j}}{\sqrt{\sigma_{x_{i}}^2 + \sigma_{x_{j}}^2}}      \label{myeqn11b}
    \end{equation}
    
    \begin{figure*}
        \centering
        \includegraphics[width=\textwidth]
        {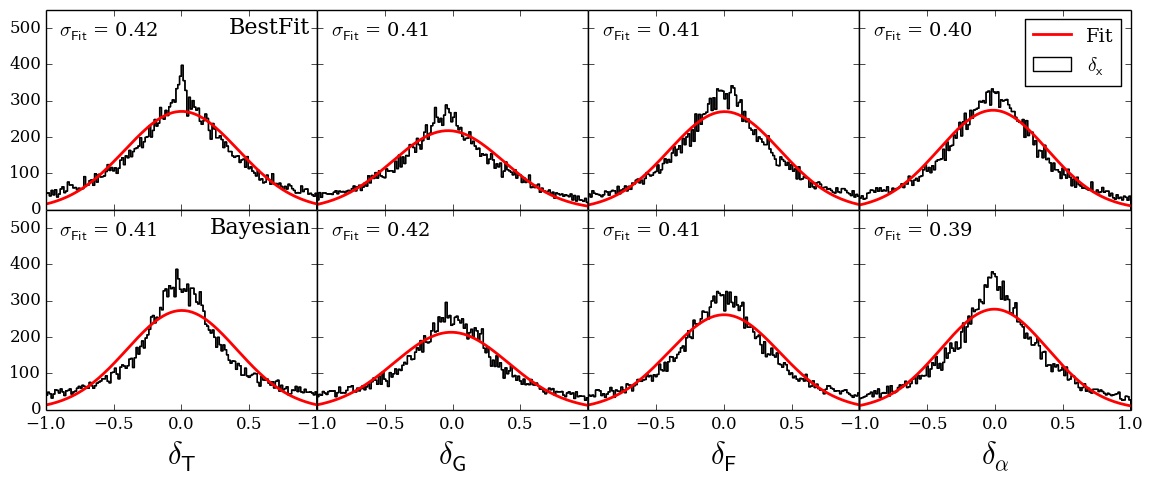}
        \caption{Histograms showing the distribution of $\delta_{x_{ij}}$ for parameters $T_{\rm eff}$, $\log g$, [Fe/H], and [$\alpha$/Fe] (abbreviated as T, G, F, and $\alpha$ for simplicity). $\delta_{x_{ij}}$ is our preferred metric for evaluating the internal consistency of our error estimates. $\delta_{x_{ij}}$ values corresponding to the BestFit parameters are shown in the top row, and those corresponding to the Bayesian parameters are shown in the bottom row. Cases in which the BestFit parameters fell on the edge of the model grid were excluded from both sets to avoid exaggerating the degree of consistency between visits.}
        \label{fig:internal_consistency_fig}
    \end{figure*}
    
    \noindent{$\delta_{x_{ij}}$ was calculated for each unique pair of distinct MaStar spectra, $i$ and $j$, that were associated with the same target. Targets with only one visit were excluded in this analysis.}
    
    Figure \ref{fig:internal_consistency_fig} shows a set of histograms representing the distribution of $\delta_{x_{ij}}$ for the BestFit and Bayesian parameters, with a Gaussian fit shown for each distribution in red. Cases in which the BestFit parameters were determined to fall on the edge of the model grid were excluded from both parameter sets to avoid giving an artificially inflated measure of their consistency. According to Equation \eqref{myeqn11b}, a standard deviation of 1 in the Gaussian fits for all of these distributions would indicate perfect consistency between the error estimates produced by our MCMC algorithm and the scatter between repeated observations. Likewise, a standard deviation of less than 1 would imply that our MCMC-based errors are overestimated by a factor of $1/\sigma_{\rm Fit}$. Therefore, according to this metric, the errors provided by the MCMC for both of our sets of parameters were overestimated by an approximate factor of $2.5$.

\subsection{External comparison}
\label{external_comparison}

    For comparison of our parameter estimates with external data sets, the APOGEE-2 catalog is the most appropriate choice. Since APOGEE-2 is a stellar survey covering near-infrared wavelengths, they use an entirely different approach to parameter determination in their ASPCAP pipeline \citep{garciaperez2016}. This makes it a good source for independent verification for some of our most critical targets. Additionally, since APOGEE has a much higher spectral resolution than MaStar, with $R\sim22,500$, we consider its parameters to be very accurate. It should be noted, however, that APOGEE primarily targeted low-temperature stars on the cool main sequence and red giant branch. This means that we have to rely on other forms of verification for hotter regions of parameter space, such as the hot main sequence and the horizontal branch.
    
    \begin{figure*}
        \centering
        \includegraphics[width=\textwidth]
        {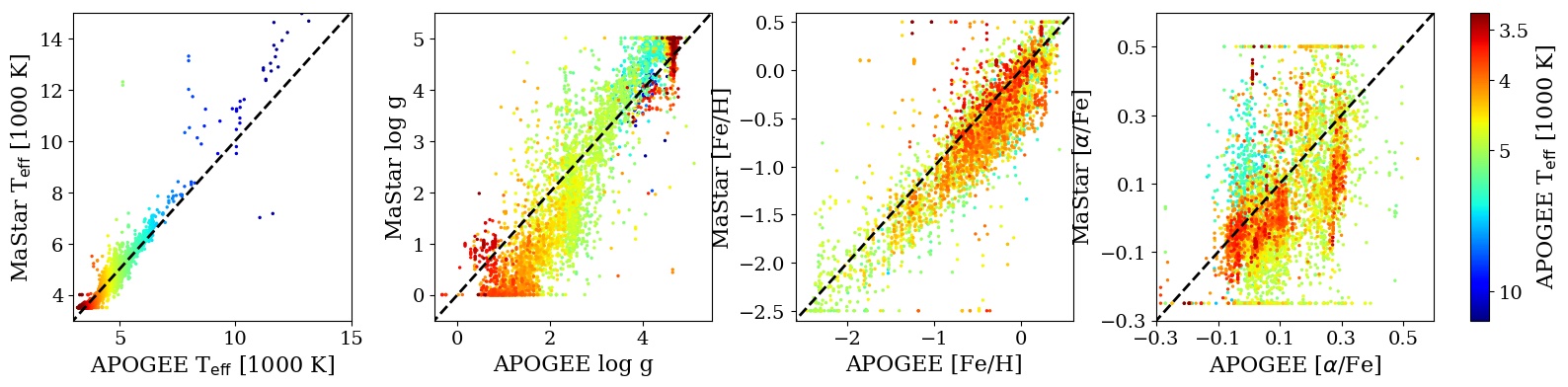}
        \caption{Direct comparison between APOGEE DR17 ASPCAP parameters and MaStar parameters.}
        \label{fig:apogee_comparison}
    \end{figure*}    
    
    Figure \ref{fig:apogee_comparison} shows a direct comparison between APOGEE-2 stellar parameters and the MaStar parameters generated using the approach described in this work. The agreement in temperature is generally quite good, with slight systematics across the temperature range crossing the one-to-one line near $5,000$ K. Above $\sim 8,000$ K, temperature measurements show less agreement for some hot stars.
    
    Comparison in $\log g$ shows good agreement at high surface gravity with a slight positive offset, and a slightly larger offset at surface gravities below $\log g = 2.5$. A clump of stars with significant scatter is visible near the center of the distribution, at $\log g = 2.5$ according to APOGEE. This is related to our decision to de-weight wavelengths below $7800{\rm \AA}$, which limited the influence of certain spectral features that are useful for constraining $\log g$ for hot stars. This was a calculated trade-off made in favor of obtaining better performance for the cool main sequence, and the improvement we saw from this is partially demonstrated in Figure \ref{fig:unweighted_v_weighted_fig}. Figure \ref{fig:logg_feh_weighting_comparison} further demonstrates this improvement by showing that the artificial feature present at the high-$\log g$ end was eliminated by the wavelength weighting entirely. Performance was also improved for warmer stars with $\log g > 3.0$, where the offset was slightly reduced.
    
    Our agreement with APOGEE in [Fe/H] is also generally good, though some nonlinear systematics may be present. This mostly affects parameters near the middle of our metallicity range. Referring back to Figure \ref{fig:logg_feh_weighting_comparison}, we see that our [Fe/H] estimates also benefited substantially after implementing wavelength weighting, especially for cool stars. Metallicity scatter may have increased slightly for hotter stars, but the effect of this is mild.
    
    According to the far-right panel of Figure \ref{fig:apogee_comparison}, our agreement with APOGEE in $\alpha$ abundance is also reasonable. This is especially clear when considering the relatively narrow range in this parameter dimension. The APOGEE parameters have more distinct high and low $\alpha$ abundance populations, whereas our distribution shows some overlap between the two. This may suggest that our routine had some difficulty constraining [$\alpha$/Fe], leading to greater scatter. This is not surprising for hot stars, since spectral features sensitive to [$\alpha$/Fe] disappear at higher temperatures. This explains why we see the most scatter in stars around and above $5,000$ K. The fifth case discussed in Appendix \ref{mcmc_behavior} demonstrates this effect. This is also the most likely explanation for some of the artifacts visible in Figure \ref{fig:alpha_raw_params}. In the right panel, showing our [$\alpha$/Fe] estimates computed using the Bayesian approach, many of our hottest stars appear to have accumulated around [$\alpha$/Fe]$=0.1$. The absence of this artifact in the BestFit estimates indicates that the MCMC step chains for these stars tended to be spread out over a wide range in [$\alpha$/Fe].
    
    \begin{figure*}
        \centering
        \includegraphics[width=\textwidth]
        {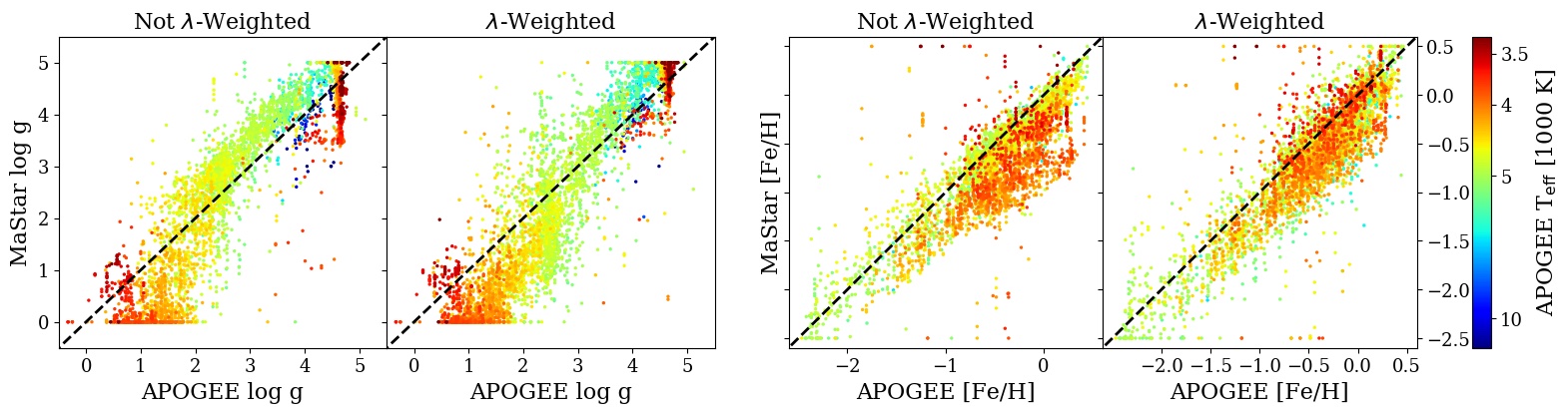}
        \caption{Direct comparison with APOGEE DR17 ASPCAP parameters shown with and without wavelength weighting implemented for $\log g$ (left) and [Fe/H] (right).}
        \label{fig:logg_feh_weighting_comparison}
    \end{figure*}        

    Another helpful set of comparison data is provided with Gaia DR2 \citep{gaia_2018}. Gaia provides accurate parallax and photometric measurements in the G, G$_{\rm BP}$, and G$_{\rm RP}$ bands. Combining the photometry with the Gaia-based distance estimates from \cite{Bailer_Jones_2018}, and the Bayestar19 3D dust map produced by \cite{Green_2019}, one can derive the extinction-corrected absolute G-band magnitude and $G_{\rm BP}-G_{\rm RP}$ color (Yan et al. in prep). This allowed us to construct a color-magnitude diagram (CMD) analogous to our $\log g$ versus $T_{\rm eff}$ diagram, shown in Figure \ref{fig:gaia_HR_diagrams_grid}. It is important to note that the Gaia CMD is not a perfect one-to-one analog to the $\log g$ versus $T_{\rm eff}$ diagram as magnitude and color are not always good proxies for surface gravity and temperature. Given this, the CMD serves primarily as a guide for the broader regions of parameter space, rather than a data source for direct one-to-one comparison similar to APOGEE.
    
    The CMDs shown in Figure \ref{fig:gaia_HR_diagrams_grid} are color-coded by each of our four atmospheric parameters. The top-left panel shows that our temperature estimates are consistent with Gaia's $G_{\rm BP}-G_{\rm RP}$ measurements, given the rough correlation between the two. This relationship becomes degenerate, however, where $G_{\rm BP}-G_{\rm RP} \lesssim 0$. This is where the upper main sequence and the horizontal branch overlap in the CMD, as opposed to the $\log g$ versus $T_{\rm eff}$ diagram, in which they differ in $\log g$. Some extreme-horizontal branch stars and white dwarfs are also present in the CMD as well, showing up at higher M$_{g}$ than the body of the main sequence.
    
    The top-right panel in Figure \ref{fig:gaia_HR_diagrams_grid} is color-coded by $\log g$. This also shows consistency with Gaia color and magnitude measurements over the entire parameter space, except for the tip of the red giant branch. Here, a small group of targets have been assigned higher $\log g$ values than the Gaia data would suggest. This is due to a combination of factors that we have already discussed. The fact that our model set cuts off abruptly at $\log g = 0$ forces some red giant branch tip stars to be matched with higher surface gravity models. This is a good example of how the Gaia data can be used to identify such cases.
    
    The bottom-left panel of Figure \ref{fig:gaia_HR_diagrams_grid} is color-coded by [Fe/H]. The most interesting thing to note here is the metallicity gradients visible in the mid giant branch and cool main sequence. This is expected, and suggests good consistency between our [Fe/H] measurements and the Gaia color data throughout most of parameter space. The hotter region (bluer on the CMD) appears to display more scatter, but this is partially due to the aforementioned overlapping between the upper main sequence and the horizontal branch. 
    
    The bottom-right panel of Figure \ref{fig:gaia_HR_diagrams_grid} is color-coded by [$\alpha$/Fe]. Similar to the previous panel color-coded by overall metallicity, abundance gradients are visible here in the mid giant branch and the cool main sequence. The red giant branch tip also shows an anomalously high $\alpha$ abundance, similar to what we saw in the panel color-coded by $\log g$. This is simply another manifestation of the same mismatch effect due to model limitations that we described previously.
    
\begin{figure*}
    \centering
    \includegraphics[width=\textwidth]
    {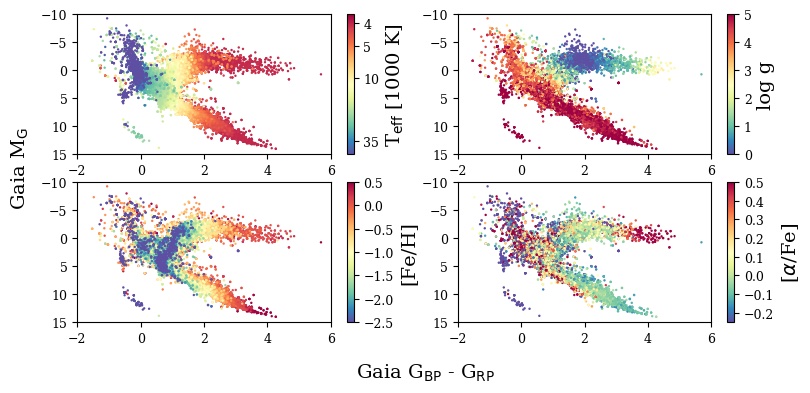}
    \caption{Gaia color-magnitude diagrams color-coded by BestFit parameter estimates. The temperature panel (top left) has had its color-coding scaled to make the temperature gradient among cooler stars more visible.}
    \label{fig:gaia_HR_diagrams_grid}
\end{figure*}

\subsection{Quality control}
\label{quality_control}

    In the final release of this parameter set, we have included several quality control flags, which fall into two categories.

    \begin{enumerate}
        \item \textbf{Low-Temperature Artifact}: Points that were placed between the cool main sequence and the upper red giant branch at temperatures near the lower limit of the model grid. These were selected by taking spectra with $T_{\rm eff} <$ $3540$ K and $0.5 < \log g < 4.0$.
        
        \item \textbf{[$\alpha$/Fe] Edge Cases}: Points that were assigned [$\alpha$/Fe] values lying at the very top or bottom of the model grid's range. This indicates our routine's failure to constrain $\alpha$ abundance in these cases. These were selected by taking spectra with [$\alpha$/Fe] $ < -0.22$ or [$\alpha$/Fe] $ > +0.47$.
        
    \end{enumerate}
    
    The final parameter space distributions after implementing these flags are shown in Figure \ref{fig:params_QC}. The important distinction between these categories is that, for Category 1, all stellar parameters have been flagged, and their use is not generally recommended. Category 2, on the other hand, has been addressed by a flag applied to the [$\alpha$/Fe] parameters only. That is, despite our considering the $\alpha$ abundance values to be invalid in these cases, we consider the remaining atmospheric parameters to still be trustworthy.
    
    The data points that make up Category 1 are easily visible in Figure \ref{fig:bestfit_raw_params}, in both the spur extending upward in $\log g$ from the red giant branch tip and downward in $\log g$ from the cool main sequence. Our decision to recommend omitting these data points is largely based on the fact that real stars generally do not exist in that part of parameter space. That is to say we believe these features to be part of the same artifact, which is related to model inaccuracy at low $T_{\rm eff}$. Omitting these points also removes the near-vertical linear structures in the upper-right corner of the left panel of Figure \ref{fig:alpha_raw_params}, further implying the artificial nature of these features.
    
    The data points in Category 2 are visible in Figure \ref{fig:alpha_raw_params}, at the top and bottom of the left panel. These are cases in which the MCMC had excessive difficulty constraining [$\alpha$/Fe].

    \begin{figure*}
        \centering
        \includegraphics[width=\textwidth]
        {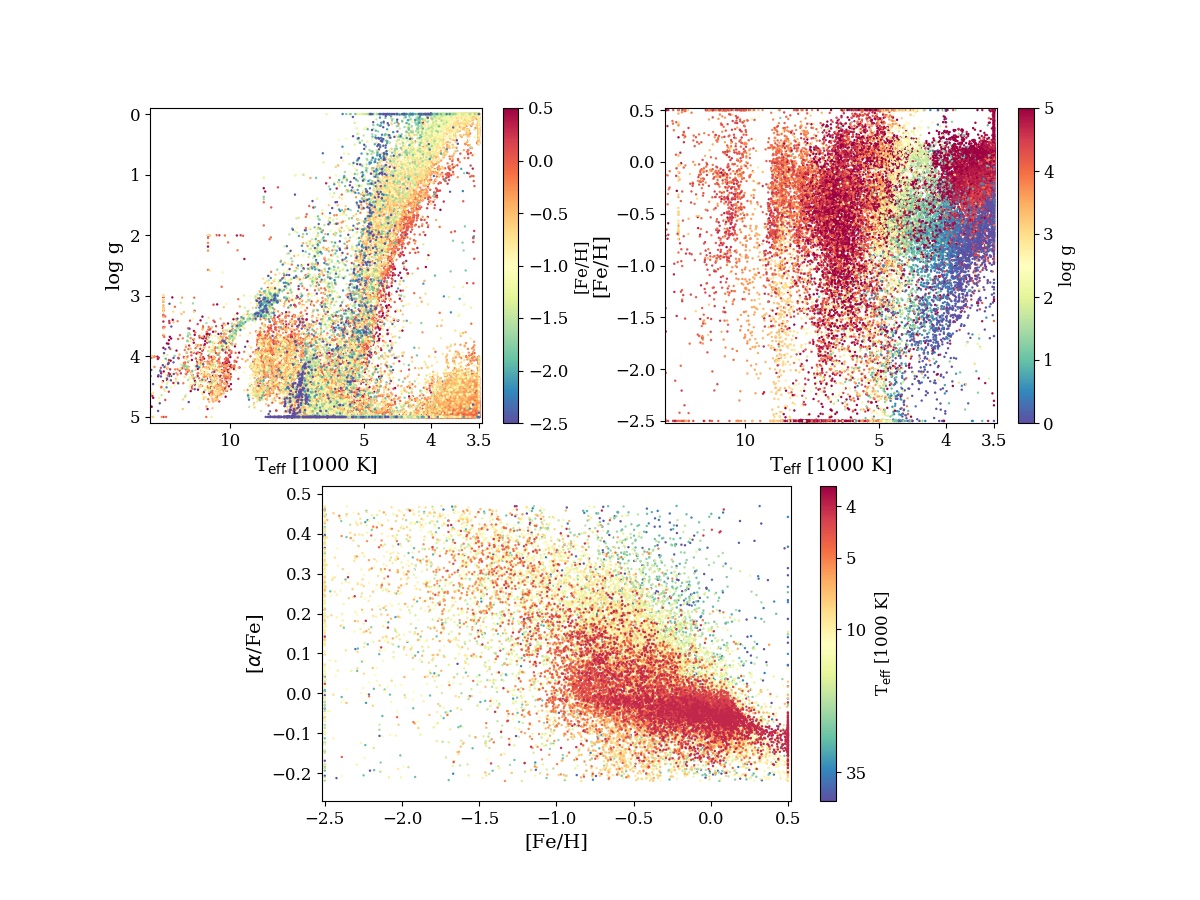}
        \caption{Distribution of MPL-11 BestFit parameters with flagged entries omitted.}
        \label{fig:params_QC}
    \end{figure*}

\subsection{Accuracy of extinction estimates}
\label{extinction_comparison}

    As we discussed in section \ref{extinction_fitting}, we believe our extinction measurements to be at least as reliable for warm and hot stars ($T_{\rm eff} \gtrsim 5,000$ K) as those calculated from the most up-to-date 3D dust map data using the Bailer-Jones distance estimates from Gaia. A direct comparison between our A$_{\rm V}$ estimates and reddening estimates given by 3D dust map data is shown in Figure \ref{fig:extinction_comparison_nocut}. It is important to note that the 3D-dust-map-based reddening is given by the arbitrary reddening coefficient $E$ (as opposed to $E(B-V)$, for example). This comparison shows a strong linear correlation for a large percentage of targets. Performing a linear fit on this distribution gives a slope of $3.37$, though this is partially affected by outliers. For a number of stars that the 3D dust map suggests have near-zero A$_{\rm V}$, our approach suggests much higher extinction. We can better understand this disagreement by looking at both the temperature and standard A$_{\rm V}$ error associated with these targets.
    
    \begin{figure}
        \centering
        \includegraphics[width=\columnwidth]
        {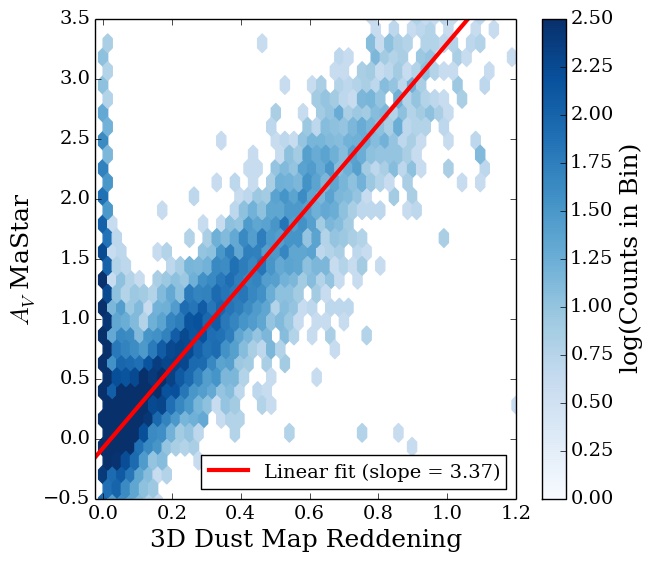}
        \caption{Distribution of MaStar A$_{\rm V}$ estimates compared with reddening measurements prior to quality cuts. Reddening estimates were made using recent 3D dust maps combined with Bailer-Jones distance estimates, and are given as the arbitrary reddening coefficient $E$. The corresponding linear fit is shown in red}
        \label{fig:extinction_comparison_nocut}
    \end{figure}
    
    Figure \ref{fig:extinction_error_fig} shows our A$_{\rm V}$ uncertainty estimates as a function of A$_{\rm V}$, color-coded by temperature. Here, it is clear that the error estimates for many cool stars show a strong positive correlation with A$_{\rm V}$ itself. These problematic A$_{\rm V}$ estimates result from molecular bands interfering with the linear fitting described in Section \ref{extinction_fitting}. These points can be easily excluded by rejecting entries for which the formal A$_{\rm V}$ error exceeds $0.005$ mag.
    
    In addition, it should be noted that in a number of cases, our method yielded a negative slope in the linear fit described in Figure \ref{fig:extinction_fig}. This corresponds to a negative A$_{\rm V}$, which is nonphysical. However, we found that, for the majority of such cases, the degree to which A$_{\rm V}$ was found to be negative is consistent with the expected uncertainty in flux calibration. This issue affects approximately $8,000$ spectra, whose negative A$_{\rm V}$ estimates have an root-mean-square deviation from 0 of $\sigma = 0.138$ mag. To eliminate outlying cases, we flagged the spectra for which $A_{\rm V} < -3\sigma$ from the final set. This left approximately $7,000$ spectra for which A$_{\rm V}$ was permitted to be slightly negative.
    
    Figure \ref{fig:extinction_comparison_fig} shows the same direct comparison between our A$_{\rm V}$ estimates and those from the 3D dust map data after applying the $0.005$ mag error cut and removing the flagged negative A$_{\rm V}$ values. This yields a much more uniform agreement at low extinction than we saw before. Performing a new linear fit on this distribution gives the following empirical relationship between A$_{\rm V}$ and the reddening coefficient, $E_{Dust Map}$:
    
    \begin{equation}
    A_{\rm V} = 3.31 E_{\rm Dust Map} - 0.076.
    \end{equation}
    
    \noindent{This relationship is consistent with the one derived by Green et al. between $E(g-r)$ and the independently derived A$_{\rm V}$ estimates that they used for testing.}
    
    \begin{figure}
        \centering
        \includegraphics[width=\columnwidth]
        {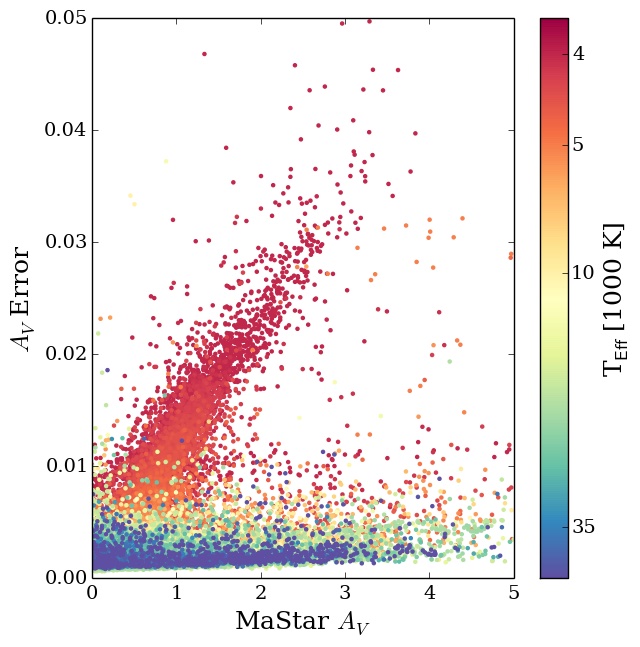}
        \caption{A$_{\rm V}$ errors plotted as a function of A$_{\rm V}$ estimates, color-coded by $T_{\rm eff}$.}
        \label{fig:extinction_error_fig}
    \end{figure}
    
    \begin{figure}
        \centering
        \includegraphics[width=\columnwidth]
        {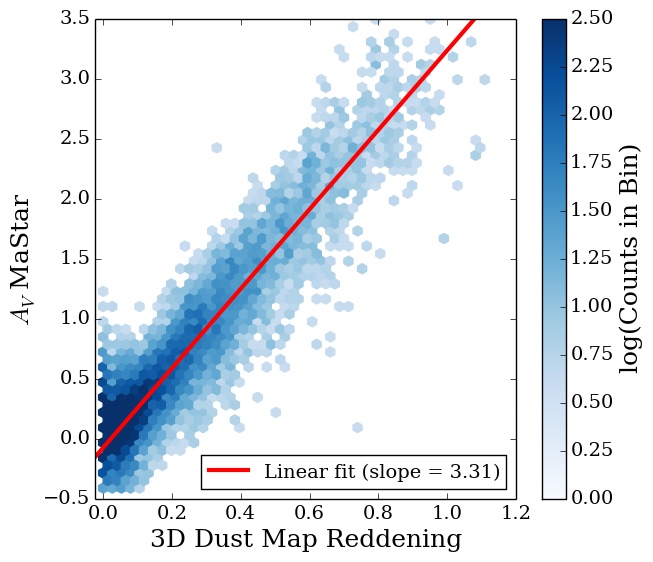}
        \caption{Distribution of MaStar A$_{\rm V}$ estimates compared with reddening measurements based on 3D dust maps following quality cuts. Spectra with $A_{V, error} > 0.005$ have been excluded from this distribution. The new corresponding linear fit is shown in red.}
        \label{fig:extinction_comparison_fig}
    \end{figure}
    
    The uncertainties in A$_{\rm V}$ included with our parameter set were given by the standard error on the linear fit described in Section \ref{extinction_fitting}. Due to flux-calibration systematics, we believe these uncertainties to be significantly underestimated. To get a better estimate of typical uncertainties in A$_{\rm V}$, we can use repeated observations to calculate a median absolute deviation (MAD) in A$_{\rm V}$ associated with each target. The MAD can then be normalized to have the same metric as $\sigma$ by dividing by $0.6745$ \citep{beers1990}, giving a quantity that is more representative of the true uncertainty. Doing this suggests that stars with $T_{\rm eff} \gtrsim 5,000$ K have a typical true uncertainty in A$_{\rm V}$ of $0.049$ mag. To evaluate how this compares with uncertainties associated with 3D Dust Map estimates, we analyzed the residuals of our A$_{\rm V}$ versus E$_{\rm Dust Map}$ comparison. This gave a median scatter of $\sigma_{Resid} = 0.205$. We expect the relative contributions to this scatter from both the MaStar data and the 3D Dust Map data to contribute in the form of a quadratic sum, such that:
    
    \begin{equation}
    \sigma_{Resid} = \sqrt{(3.31)^2\sigma_{E\_\rm Dustmap}^2 + \sigma_{Av\_\rm MaStar}^2}.
    \end{equation}
    
    \noindent{Taking $\sigma_{Av\_{\rm MaStar}} = 0.049$ and solving this equation gives $\sigma_{E\_{\rm Dustmap}} = 0.06$. This estimate of the typical uncertainty on E$_{\rm Dust Map}$ is roughly consistent with the systematic error floor of $0.08$ that Green et al. placed on the uncertainty in $E(g-r)$ \cite[Figure~16]{Green_2019}. With the slope of $3.31$, the uncertainty from the 3D dust map measurements contributes $0.20$ mag to the vertical scatter in Figure \ref{fig:extinction_comparison_fig}, which is much larger than the contribution by the uncertainty of our MaStar-based extinction estimates. Thus, we conclude that our A$_{\rm V}$ estimates are superior in precision.}
    
\subsection{Inaccuracies in the model spectra}
\label{inaccuracies_in_the_model_spectra}
    
    One promising application of large stellar libraries similar to MaStar lies with the large-scale comparison between empirical spectral data with present cutting-edge models. While our parameter-measurement efforts have placed the most emphasis on the cases in which the model fitting performs as desired, cases in which it has performed in unexpected ways can be just as informative. For example, the previously mentioned gap in the hot main sequence at approximately $9,250$ K raises interesting questions about the BOSZ models and their representation of certain spectral features in this parameter regime. For stars near this temperature, the most obvious spectral features are the Balmer series of absorption lines. We suspect that the cause of this gap is that the equivalent widths of Balmer lines reach their maximum around this temperature. If the model spectra do not have the same maximum equivalent width (EW), it would result in a bias in temperature measurements. To further test this statement we plotted EW as a function of color.
    
    We measured the synthetic colors for both the data and models using the response curves made available by the Gaia team with Gaia DR2. We then plotted the colors of the observed spectra against the equivalent widths of the first four Balmer absorption lines measured in them and their corresponding best-fit models. This is shown in the four panels on the right side of Figure \ref{fig:EW_color_balmer_lines}. The line profile mismatch is relatively mild in the case of H$\alpha$. For the three subsequent Balmer lines, however, the disagreement is much stronger. This is even better illustrated if we look at the behavior of the equivalent widths in the models as is shown in the four panels on the left side of Figure \ref{fig:EW_color_balmer_lines}, represented by solid lines corresponding to the $\log g$ bins of the model grid. For H$\beta$ and H$\gamma$, the models corresponding to the highest $\log g$ bins ($4.4$-$5.0$) extend far above the equivalent widths given by the data, except for a few outliers. This implies that, in order for the models with the correct $T_{\rm eff}$ and $\log g$ to be selected, the MaStar spectra would have had to have much larger H$\beta$ and H$\gamma$ equivalent widths. To resolve this disagreement, the algorithm tended to select $T_{\rm eff}$ to be either hotter or cooler by $200-300$ K, leading to a gap in the final distribution. An example of typical behavior of the MCMC around the gap is shown in the sixth case discussed in Appendix \ref{mcmc_behavior}, in which two local minima in the MCMC sample appear to be present on either side of the gap.
    
    Several specific examples of this equivalent width discrepancy are shown in Figure \ref{fig:Hgamma_examples_fig}. Here, we have compared H$\gamma$ lines from three MaStar spectra with models generated in two different ways. The first set of models (shown in red) correspond to the BestFit parameters given by our routine. The alternate models (shown in magenta) correspond to entries in the MCMC step chains that agree closely with temperatures estimated based on Gaia data using a color-temperature relationship described in \cite{mucciarelli2020gaia}. These are meant to give a reasonable estimate of where in parameter space these stars ought to lie. What we see from this is that the alternate models provide a considerably worse match to the H$\gamma$ line profile present in the data, whereas the BestFit models provide a near-perfect match.
    
    The implication of these findings is that the Balmer series equivalent widths in the models disagree with those present in the data. This suggests some fundamental inaccuracy in the BOSZ models for these particular features. This problem could have been caused by the spectral synthesis code used for the BOSZ models, \textsc{SYNTHE}, handling radiation damping incorrectly for the Balmer lines (Szabolcs Mészáros, private communication). This problem could be addressed in the future by employing a newer spectral synthesis program when constructing the next generation of the models. This is an excellent example of the way stellar libraries similar to MaStar can be used to improve current theoretical spectra.

    \begin{figure*}[bp!]
        \centering
        \includegraphics[width=\textwidth]
        {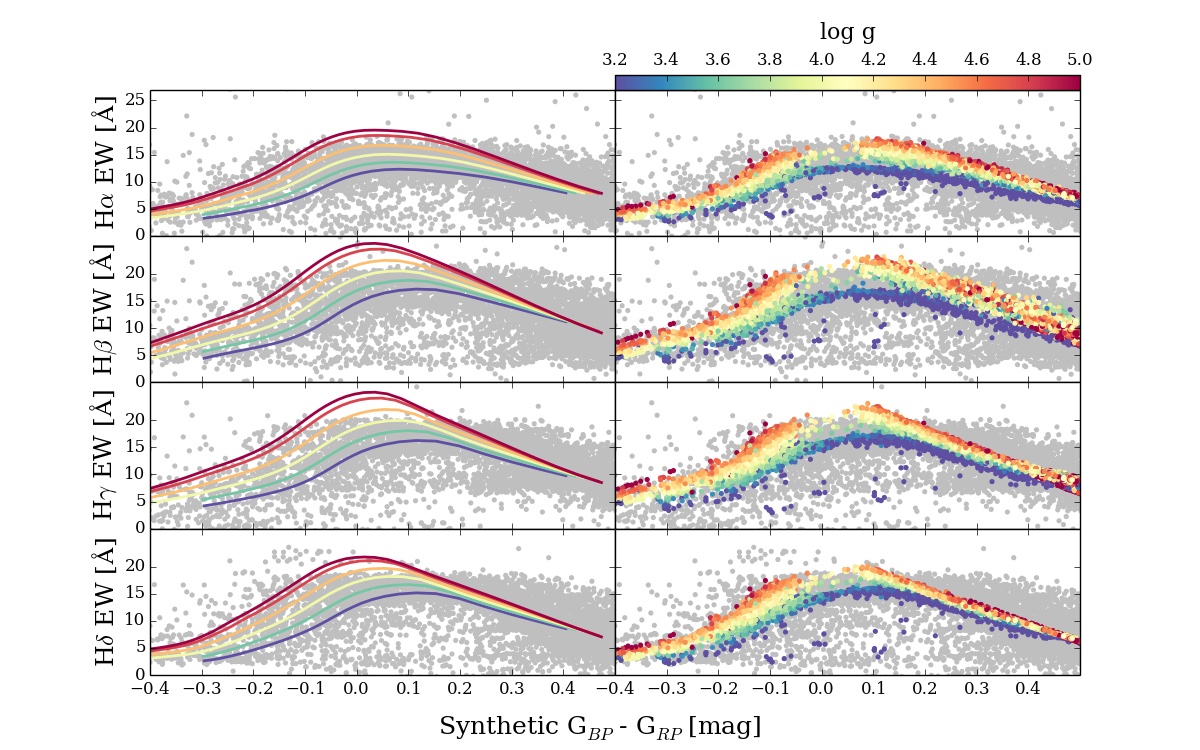}
        \caption{The relationship between the equivalent width (EW) of the first four Balmer lines and Gaia $G_{\rm BP}-G_{\rm RP}$ color for both the original model grid (left, represented by solid curves) and the interpolated BestFit models corresponding to each MaStar spectrum (right, represented by points color-coded by $\log g$). The EW-color distribution of the set of MaStar spectra, with extinction effects removed, is shown in the background of each panel in gray.}
        \label{fig:EW_color_balmer_lines}
    \end{figure*}

    \begin{figure*}[bp!]
        \centering
        \includegraphics[width=\textwidth]
        {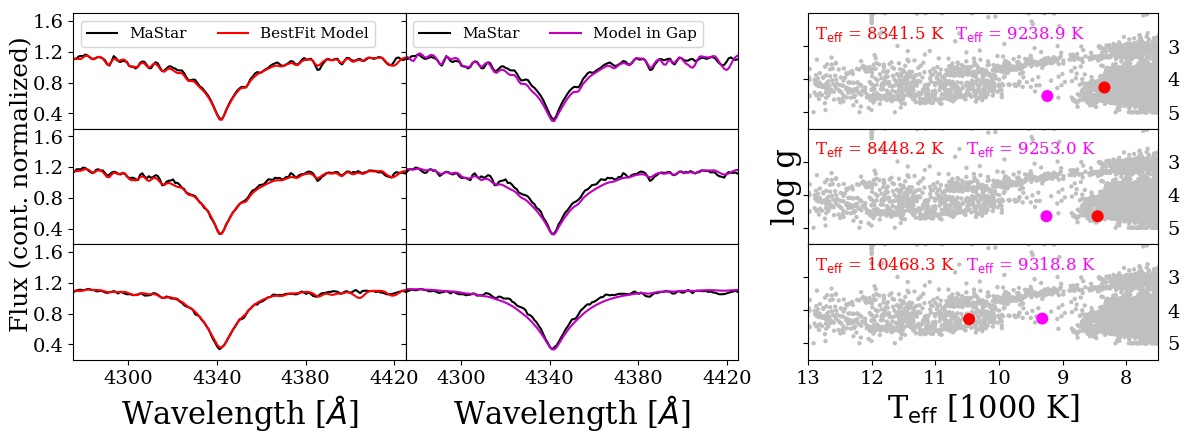}
        \caption{Several examples showing the performance of our narrow-band fitting of H$\gamma$ for the BestFit parameter solution (shown in red) compared with the H$\gamma$ line profile of an alternate model generated using a temperature computed based on Gaia $G_{\rm BP}-G_{\rm RP}$ color (shown in magenta). Both sets of model spectra are compared with the original MaStar spectrum shown in black. The relative position in parameter space of each model is shown in the three panels on the right.}
        \label{fig:Hgamma_examples_fig}
    \end{figure*}

\section{Conclusions}
\label{conclusions}
    
    In this paper, we present a stellar parameter determination method and applied it to the spectra from the MaStar library. The method has been shown to have several key characteristics and advantages. It was based purely on theoretical models without the need for priors based on external data, such as isochrones or color measurements. It combined the use of continuum-normalized full-spectrum fitting with continuum fitting, thus taking advantage of both small-scale and large-scale features. Comparison with external data from APOGEE and Gaia have shown that the method performed well over a very wide range of stellar parameter space. Stellar population models based on MaStar's first data release have been presented in \cite{maraston2020}, and these models will soon be updated using the parameters presented in this work.
    
    While our procedure's overall performance in all four parameter dimensions was good, several artifacts still show up in the final parameter distribution. Modifying our fitting routine ($\chi^{2}_{\rm LF}$ scaling, continuum rejection, and scaling the per-pixel $\chi^{2}$ contribution by wavelength), have helped to alleviate some of these artifacts, but have not been able to get rid of them completely. Our method's dependence on theoretical spectra alone helped to trace the origins of these effects back to inaccuracies in the models. While the ATLAS9-based BOSZ models have proven very accurate in most regions of parameter space, the data suggest there are at least two regions with considerable room for improvement: the warm main sequence at temperatures around $9,250$ K, and at very low temperatures, both at low surface gravity on the tip of the red giant branch, and at high surface gravity on the cool main sequence. We have shown that the $9,250$ K gap is due to the BOSZ models having too high of an equivalent width for the Balmer lines when they reach their peaks around this temperature. This could mean there are also systematic biases in the $T_{\rm eff}$ derived based on BOSZ models above and below this temperature. The MaStar spectra could be very useful for correcting the atomic and molecular line lists to improve the accuracy of the model spectra. The MaStar parameter catalog containing our BestFit results is available on the SDSS-IV DR17 website as part of version 2 of the MaStar stellar parameter value-added catalog.\footnote{\footurl}

\begin{acknowledgements}

    DL, RY, and RW would like to acknowledge support by the NSF grant AST-1715898. RY would like to acknowledge support by the Hong Kong Global STEM Scholar scheme, by the Hong Kong Jockey Club through the JC STEM Lab of Astronomical Instrumentation program, by the Direct Grant of CUHK Faculty of Science, and by the Research Grant Council of the Hong Kong Special Administrative Region, China (Projects No. 14302522). 
    
    Timothy C. Beers acknowledges partial support for this work from grant PHY 14-30152; Physics Frontier Center/JINA Center for the Evolution of the Elements (JINA-CEE), awarded by the US National Science Foundation.

    Funding for the Sloan Digital Sky Survey IV has been provided by the Alfred P. Sloan Foundation, the U.S. Department of Energy Office of Science, and the Participating Institutions. SDSS acknowledges support and resources from the Center for High-Performance Computing at the University of Utah. The SDSS web site is www.sdss.org.

    SDSS is managed by the Astrophysical Research Consortium for the Participating Institutions of the SDSS Collaboration including the Brazilian Participation Group, the Carnegie Institution for Science, Carnegie Mellon University, Center for Astrophysics | Harvard \& Smithsonian (CfA), the Chilean Participation Group, the French Participation Group, Instituto de Astrofísica de Canarias, The Johns Hopkins University, Kavli Institute for the Physics and Mathematics of the Universe (IPMU) / University of Tokyo, the Korean Participation Group, Lawrence Berkeley National Laboratory, Leibniz Institut für Astrophysik Potsdam (AIP), Max-Planck-Institut für Astronomie (MPIA Heidelberg), Max-Planck-Institut für Astrophysik (MPA Garching), Max-Planck-Institut für Extraterrestrische Physik (MPE), National Astronomical Observatories of China, New Mexico State University, New York University, University of Notre Dame, Observatório Nacional / MCTI, The Ohio State University, Pennsylvania State University, Shanghai Astronomical Observatory, United Kingdom Participation Group, Universidad Nacional Autónoma de México, University of Arizona, University of Colorado Boulder, University of Oxford, University of Portsmouth, University of Utah, University of Virginia, University of Washington, University of Wisconsin, Vanderbilt University, and Yale University.

\end{acknowledgements}

\bibliographystyle{aa}
\bibliography{aanda}

\begin{appendix} 
\section{MCMC behavior}
\label{mcmc_behavior}

    Figure \ref{fig:mcmc_convergence} shows several examples of MCMC step chains selected to illustrate the MCMC's performance in various regions of parameter space. Each row displays the step chain compiled for a given MaStar spectrum plotted in 4D parameter space using two panels. From top to bottom, the examples shown are a lower-mid RGB star, a warm MS star, a metal-poor giant star, a cool dwarf, a very hot MS star, and a moderately hot MS star falling near the $9,250$ K MS gap that is discussed in section \ref{inaccuracies_in_the_model_spectra}. The first three examples all exhibit excellent constraint in all four parameter dimensions. The fourth case (the cool M-dwarf), falling very close to the edge of the model grid in both $T_{\rm eff}$ and $\log g$, exhibits the edge-clipping effect discussed in section \ref{raw_parameter_distribution_and_uncertainty_estimates}, despite the exclusion of the higher-$\chi^{2}$ entries in the step chain. The fifth case (the very hot MS star) appears well constrained in $T_{\rm eff}$-$\log g$ space, but $[Fe/H]$ and $[\alpha/Fe]$ are essentially unconstrained. This is due to fact that spectral features lose their dependencies on metal abundances at higher temperatures. In the final case, we show an example of a star whose MCMC step chain contains two distinguishable local minima lying on either side of the $9,250$ K gap, although one was clearly preferred by the algorithm. Similar to the previous case, $[Fe/H]$ and $[\alpha/Fe]$ are completely unconstrained in this regime.

    \begin{figure*}
        \centering
        \includegraphics[width=\textwidth]
        {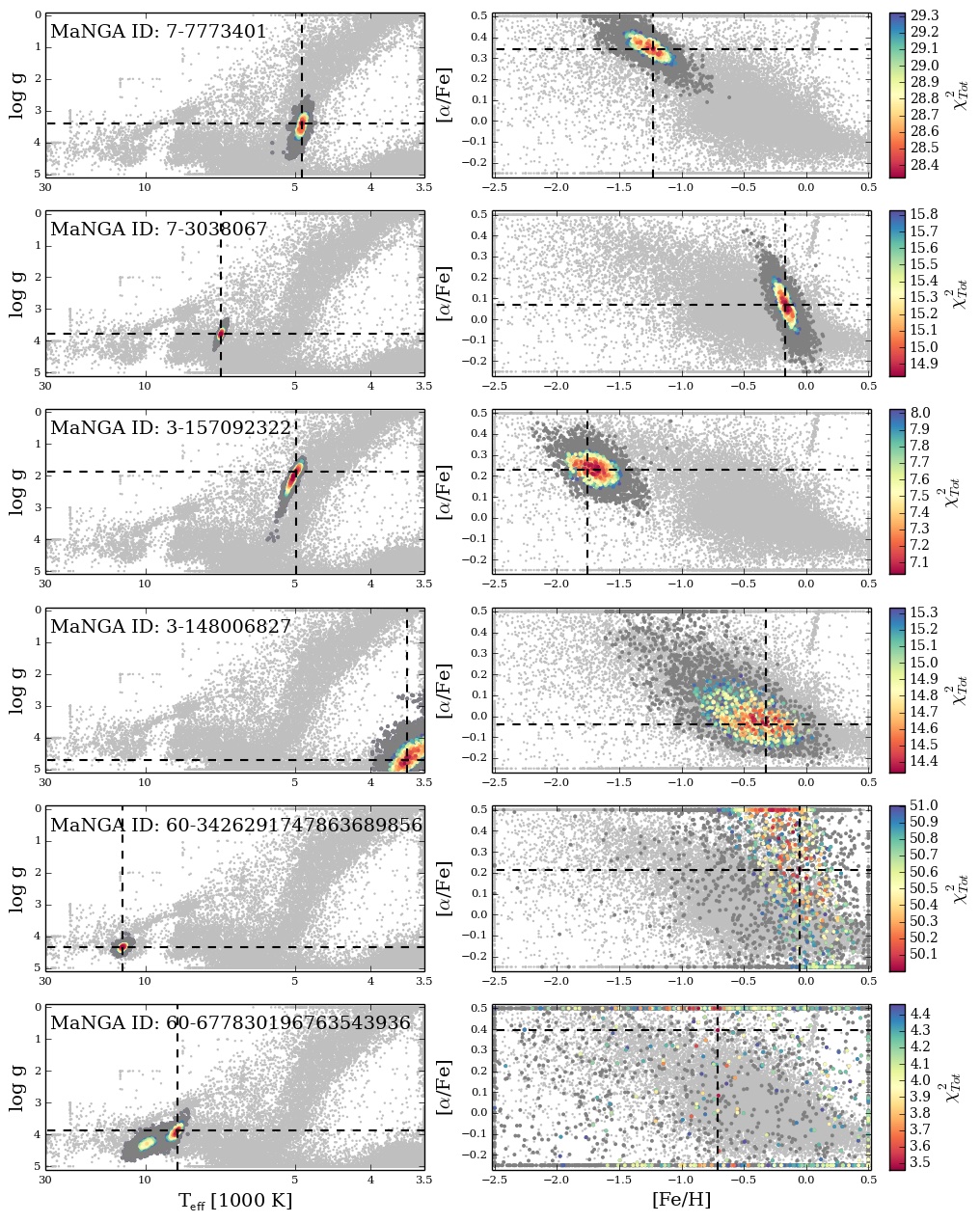}
        \caption{Several examples of MCMC step chain convergence in 4D parameter space. The portions of the MCMC step chains used for parameter and error calculations are color-coded by $\chi^{2}_{Tot}$, and the trimmed portions, consisting of the burn-in period and all points with $\chi^{2}_{Tot}$ exceeding min($\chi^{2}_{Tot}$)+1.0, are displayed in gray. The positions of the final adopted parameters are indicated by the black dashed lines.}
        \label{fig:mcmc_convergence}
    \end{figure*}

\end{appendix}

\end{document}